\documentclass[fleqn,usenatbib]{mnras}

\usepackage{newtxtext,newtxmath}

\usepackage[T1]{fontenc}
\usepackage{ae,aecompl}

\usepackage{graphicx}	
\usepackage{amsmath}	
\usepackage{amssymb}	
\usepackage{float}

\hypersetup{draft} 

\title[SIS Cross Sections in the circular case]{Strong lensing cross sections for isothermal models -- I.
\\Finite source effects in the circular case}

\author[V. P. de Freitas et al.]{
Vanessa P. de Freitas,$^{1}$\thanks{E-mail: vpacheco@cbpf.br }
Martin Makler,$^{1}$
Habib S. D\'umet-Montoya$^{2}$
\\
$^{1}$Centro Brasileiro de Pesquisas F\'isicas, Rua Dr. Xavier Sigaud 150,  Rio de Janeiro, RJ, CEP 22290-180, Brazil\\
$^{2}$Universidade Federal do Rio de Janeiro - Campus Maca\'e, Rua Alo\'isio Gomes da Silva, 50,  Maca\'e, RJ, CEP 27930-560, Brazil 
}

\date{} 

\pubyear{2018}

\begin{document}
\label{firstpage}
\pagerange{\pageref{firstpage}--\pageref{lastpage}}
\maketitle

\begin{abstract}
The strong galaxy-galaxy lensing produces highly magnified and distorted images of background galaxies in the form of arcs and Einstein rings. Statistically, these effects are quantified, for example, in the number counts of highly luminous sub-millimeter galaxies and of gravitational arcs.  Two key quantities to model these statistics are the magnification and the arc cross sections. These are usually computed using either the circular infinitesimal source approximation or ray-tracing simulations for sources of finite size. In this work, we use an analytic solution for gravitational arcs to obtain these cross sections as a function of image magnification and length-to-width ratio in closed form, for finite sources. These analytical solutions provide simple interpretations to the numerical results, can be employed to test the computational codes, and can be used for fast a computation of the abundance of distant sources and arcs. In this paper, the lens is modeled by a Singular Isothermal Sphere, which is an excellent approximation to radial density profile of Early-Type galaxies, and the sources are also axisymmetric. We derive expressions for the geometrical properties of the images, such as the area and several definitions of length and width. We obtain the magnification cross section in exact form and derive a simple analytic approximation covering the arc and Einstein ring regimes. The arc cross section is obtained down to the formation of an Einstein ring and given in terms of elementary functions. Perturbative expansions of these results are worked out, showing explicitly the correction terms for finite sources.

\end{abstract}

\begin{keywords}
gravitational lensing: strong -- methods: analytical -- galaxies: general
\end{keywords}



\section{Introduction}

Gravitational arcs and Einstein rings \citep{1985ApJ...292..348S}
are highly distorted and magnified images of distant galaxies (sources) due to the light deflection produced by foreground galaxies acting as lenses.
These images may be used to probe the mass distribution in the lens galaxies \citep[e.g.][]{TreuKoopmans, Koopmans}, including substructures and tests of the Cold Dark Matter paradigm \citep[e.g.][]{Vegetti2012,Xu2015, Li2016}; to find and study high-redshift galaxies 
through the \textit{gravitational telescope} effect \citep{2016A&A...595A.100C, 2017MNRAS.465.3558N, Goobar2017,Zavala2017}; 
to constrain cosmological models
\citep{Suyu2010, Cao, Treu2016Rev}; and to test modified gravity theories \citep{2010ApJ...708..750S, 2013JHEP...10..031E}.
For a review on galaxy-scale strong lensing, see \citet{Treu}.
 
The many applications of strong lensing, in particular by galaxies, motivated the search for arcs and Einstein rings in 
\textit{Hubble Space Telescope} (HST) 
images \citep[e.g.,][]{Hogg1996,Ratnatunga1999, 2008ApJS..176...19F, Jackson2008,Marshall2009robot},
SDSS spectroscopy with HST follow-up
\citep[SLACS survey,][]{2006ApJ...638..703B,2017ApJ...851...48S} 
and in wide-field imaging surveys from the ground. 
In particular, the searches in the Canada--France--Hawaii Telescope Legacy Survey \citep{Cabanac2007,2012ApJ...749...38M,2013ApJ...777...97S,Paraficz2016,2017MNRAS.471..167J} and in the ongoing 
DES\footnote{
\citet{DarkEnergyCollab}, \url{http://www.darkenergysurvey.org/}}  \citep{Nord2016,2017ApJS..232...15D}, 
KiDS\footnote{Kilo-Degree Survey
Survey, \citet{deJong2015}, \url{http://kids.strw.leidenuniv.nl/}} \citep{2017MNRAS.472.1129P}
and
HSC\footnote{Hyper Suprime-Cam Subaru Strategic Program, \citet{2018PASJ...70S...4A}, \url{http://hsc.mtk.nao.ac.jp/ssp/}} \citep{2017arXiv170401585S}
surveys are leading to hundreds of galaxy-scale strong lenses.
These numbers are expected increase by about an order of magnitude with the close completion of DES, KiDS and HSC.
Comparable numbers are expected from the forthcoming J-PAS\footnote{Javalambre Physics of the Accelerating Universe Astrophysical Survey, \citet{Benitez2014}} project.
These numbers will increase even further in the near future, with the operation of LSST\footnote{Large Synoptic Survey Telescope, \citet{LSSTCollab}, \url{http://www.lsst.org/}} 
and 
Euclid,\footnote{\citet{Refregier2010,laureijs_euclid_2011}, \url{http://www.euclid-ec.org/}} 
which are both expected to detect $\mathcal{O}\left(10^5\right)$ systems with arcs \citep{Collett2015}.

In recent years, strongly lensed systems started to be discovered in wide-field surveys at sub-millimeter and millimeter wavelengths, such as from the South Pole Telescope \citep{2013Natur.495..344V} and the Herschel-ATLAS survey \citep{2017MNRAS.465.3558N}. In this case, the systems are not identified by their arc or ring shape, but by the large magnification of the sources (in this case dusty star-forming galaxies). The surveys carried out with Herschel are expected to deliver a sample of more than a hundred of sub-mm bright strongly lensed galaxies \citep{2017MNRAS.465.3558N}, and could reach around a thousand, depending on the detection technique
\citep{2012ApJ...755...46L}.

With such large numbers of objects from optical and (sub-)mm surveys, it becomes intractable to perform a detailed modeling of each system, in particular considering the difficulty in obtaining high-resolution imaging  and redshifts for the sources. An alternative approach, known as strong lensing statistics,  
is to calculate probability distributions of observable properties of arcs, such as their length-to-width ratio ($L/W$) and magnification, and compare them to observations \citep{1988ApJ...324L..37G, 1993ApJ...403..509M, Wu, Bartelmann, 2009A&A...508..141F, 2010MNRAS.406.2352L,2010ApJ...717L..31L}.
For an excellent review on arc statistics, see \citet{2013SSRv..177...31M}.

To compute the abundance of arcs 
as a function of their properties, one needs to know the number densities of lens and sources as a function of some of their properties, including their redshift. These can be empirically determined from observational data or be represented by families of models with parameters to be determined from the arc statistics observations. 
Strong lensing statistics has been applied to several problems, in particular involving galaxy scale lenses, such as to predict the number of arcs found in the SLACS survey \citep{Dobler2008} and to explain the abundance of highly luminous sub-millimeter galaxies  
\citep{2010MNRAS.406.2352L,2010ApJ...717L..31L,2011ApJ...734...52H,2012ApJ...755...46L}.

Another key ingredient in strong lensing statistics is
the efficiency to produce images with given properties --- such as
magnification and length-to-width ratio. 
This efficiency 
is encoded 
in the \textit{cross section} and the main aim of the paper is to compute this quantity for a specific lens and source model.

During the past decade, several studies using a diverse set of observables --- specially weak and strong lensing --- and simulations 
have shown that the radial density profile of 
galaxy-scale lenses (i.e. Early-Type galaxies)
is surprisingly close to the Singular Isothermal Sphere (SIS) profile \citep[see e.g.,][and references therein]{Gavazzi, Koopmans,  deVen, Blundell, Treu, Bolton, Grillo, 2012ApJ...755...46L, 2013ApJ...777...98S,2018MNRAS.476.4383D}, which is given by 
\citep{Turner, Binney1987, Schneider}:

\begin{equation}
\rho(r) = \dfrac{\sigma_v^2}{2\pi G}\dfrac{1}{r^2},
\label{eq:sisdensity}
\end{equation}
where $\sigma_v^2$ is the one-dimensional velocity dispersion.
Remarkably, lens models based on this solution and including external shear and/or ellipticity, allow one to derive analytic solutions for several lensing related quantities \citep[see e.g.,][D\'umet-Montoya et al., in prep.]{2005ApJ...634...77I,Dobler2006,Dobler2008,2013ApJ...770L..34C,2013MNRAS.430.1423E},
including gravitational arcs. 

Generally, two approaches have been used for computing the cross section for arc statistics:
either the source is considered infinitesimal and several calculations can be carried out analytically \citep[see e.g., ][]{2001ApJ...559..572O, 2003MNRAS.340..105M, 2013arXiv1308.6569C,Habib2013,2013MNRAS.430.1423E} or ray-tracing simulations are carried out producing images of finite sources \citep{Wu, 1993ApJ...403..509M, Bartelmann,2005APh....24..257H,2011ApJ...734...52H,2012ApJ...755...46L,2012A&A...547A..66R}.
The latter are more realistic, but also more time consuming. Furthermore, the simulations have to be carried out again for each change of parameter and the results cannot always be interpreted in a transparent way.

Here we take an alternative approach, which is to use the analytical solutions for arcs in the SIS case to derive the cross sections. In this way we are able to introduce and study the finite source effects 
in an analytic or semi-analytic way. 
Furthermore, the treatment
enables us to tackle the problem down to the formation of Einstein rings, which cannot be addressed with infinitesimal sources. 

An advantage of analytic solutions is that they offer the possibility of a more clear physical interpretation of the results. They enable to probe the whole parameter space involved and can be used for fast calculations. They can also be used to test the accuracy of numerical codes
that 
are developed for more generic models, in the specific situations where the analytical results hold. 
Therefore, there is a complementarity with fully numerical approaches and it is worth to search for such analytical solutions.

In this work we consider the simple case of a circular SIS model with circular sources (elliptical sources are addressed in a separate paper). We start by investigating the geometrical properties of the images, seeking to obtain the magnification and $L/W$ in a closed form. 
We test several definitions of length and width and apply the results to compute the cross sections of magnification and arc formation, which can be used to predict the abundance of distant sources as a function of flux 
and arcs as a function of $L/W$, respectively. 
Remarkably, 
this problem can be treated analytically all the way down to the computation of the cross sections and we are able to express them, under some approximations, in terms of elementary functions.
We also obtain perturbative solutions that explicitly show the correction terms for finite sources. From the solutions obtained, we are able to clarify some properties empirically found in more general situations using simulations.

This paper is organized as follows: in Section~\ref{sec: ArcinSIS}, we present a brief review of SIS lenses and the solution for finite sources. In Section~\ref{sec:GeometricalProperties}, we derive expressions for the magnification, length and widths of the images, which are used in Section~\ref{sec:Cross Section} to obtain the magnification and arc formation cross sections.
In Sections~\ref{sec:summary} and \ref{sec:Conclusion}, we summarize and discuss our results.
In Appendix~\ref{sec:arcellipse}, we compare the solutions for the SIS arcs to the ArcEllipse geometrical figure.
Finally, in Appendix~\ref{sec:MagxR}, we discuss the semi-analytic method introduced by \citet{Fedeli} for finite sources in the context of the arc cross sections obtained in this paper.

\section{Arcs in the SIS model}
\label{sec: ArcinSIS}

In this section, we present a brief overview of axially symmetric singular isothermal lens models 
and circular sources to fix the notation and provide the basic expressions to be used in the paper.

\subsection{Lensing by a Singular Isothermal Sphere}
\label{sec:basics}

 The lensing properties are encoded in the lens equation, which relates the position of the observed images $\mathbf{\xi}$ to those of the source $\mathbf{\eta}$. By choosing a characteristic length-scale $\xi_0$ and defining $\mathbf{x} \equiv \mathbf{\xi}/\xi_0$ and $\mathbf{y} \equiv  \mathbf{\eta}/\eta_0$, where $\eta_0 \equiv D_{OS}\xi_0/D_{OL}$, and $D_{OS}$ and $D_{OL}$ are the angular diameter distances from the observer to the source and the lens, respectively, lens equation can be written in dimensionless form  \citep{Schneider, Petters, Mollerach}:
\begin{equation}
\mathbf{y} = \mathbf{x} - \mathbf{\alpha}(\mathbf{x}),
\label{eq:yx}
\end{equation}
where $\mathbf{\alpha}(\mathbf{x})$ is the dimensionless deflection angle. 

The local distortion in the lens plane is described by the Jacobian matrix of the transformation~(\ref{eq:yx}) 
\begin{equation}
\mathbb{J} = \left(\frac{\partial \mathbf{y}}{\partial \mathbf{x}}\right)_{ij} = \delta_{ij} - \partial_i\alpha_j(\mathbf{x}). 
\label{Jaco}
\end{equation}

The eigenvalues of the Jacobian matrix give the inverse of the magnification in the tangential and radial directions and can be written as
\begin{equation}
\lambda_{r,t} (\mathbf{x}) = \mu_{r,t}^{-1}(\mathbf{x}) = 1 - \kappa(\mathbf{x}) \pm \gamma(\mathbf{x}),
\label{eq:lambdas}
\end{equation}
where $\kappa(\mathbf{x})$ and $ \gamma(\mathbf{x})$ are the convergence and the shear. The positive sign gives the eigenvalue associated to the tangential eigenvector and the negative sign corresponds to radial one. For axially symmetric lens models we have
\begin{equation}
\kappa (x) = \frac{1}{2} \left[\frac{\alpha(x)}{x} + \frac{d\alpha(x)}{dx}\right], \;
\gamma (x) = \frac{1}{2} \left[\frac{\alpha(x)}{x} - \frac{d\alpha(x)}{dx}\right],
\end{equation}
where $x = \vert \mathbf{x}\vert$ is the radial coordinate.

 The sets of points for which $\lambda_{r,t} (\mathbf{x})= 0$ determine the radial and tangential critical curves, respectively. Mapping these curves onto the source plane give us the caustics. 

For the SIS density profile (equation~(\ref{eq:sisdensity})), if we choose the length-scale $\xi_0$ as the Einstein radius
\begin{equation}
\xi_0 = R_E = \frac{\sigma_v^2}{G \Sigma_{\rm crit}},
\label{RE}
\end{equation}
where $\Sigma_{\rm crit}$ is the critical surface mass density
\begin{equation}
\Sigma_{\rm crit} = \frac{c^2}{4\pi G}\frac{D_{OS}}{D_{OL}D_{LS}},
\end{equation}
and $D_{LS}$ is the angular diameter distance from the lens to the source, then the convergence, shear and deflection angle are
\begin{equation}
\kappa (x) = \gamma (x) = \frac{1}{2x}, \; \; \; \; \; \;
\mathbf{\alpha}(x) = \hat{x},
\label{kappagamma}
\end{equation}
and the lens equation~(\ref{eq:yx}) is
\begin{equation}
\mathbf{y} = \left(x-1\right)\hat{x},
\label{eq:sislenseq}
\end{equation}
where $\hat{x}$ is the radial unit vector.

For $y =|\mathbf{y}|< 1$ this equation has two solutions, one with $ x < 1$, i.e., the image is inside the Einstein ring, and one with $ x > 1 $, which we call the internal and external images, respectively. For $y > 1$ there is only one solution and the boundary where the multiplicity changes (the curve $y=1$) is often referred to as radial pseudo-caustic \citep{Dobler2006} and we will keep this terminology along the text.   

Substituting expressions (\ref{kappagamma}) in equation~(\ref{eq:lambdas}), the eigenvalues of the Jacobian matrix, are given by
\begin{equation}
\lambda_r = 1, \  \  \ \lambda_t = 1-\frac{1}{x}.
\label{lambdasis}
\end{equation}
Therefore, the tangential critical curve is given by $x =1$, i.e., the Einstein radius, and there is no radial critical curve.

The change in shape of infinitesimal sources is given by a linear transformation defined by the Jacobian in Eq.~(\ref{Jaco}). In particular an infinitesimal circular source of radius $R_0$ will be mapped into an ellipse whose semi-axes in the tangential and radial directions will be given by, respectively $a = R_0\vert \mu_t\vert$ and $b = R_0\vert\mu_r\vert$. Therefore, the axial ratio of the image, $R_\lambda$, and the magnification, $\mu$, which is the ratio between the areas of the image and the source, will be given by
\begin{equation}
\label{eq:Rlambda}
 R_\lambda = \bigg| \dfrac{\mu_t}{\mu_r} \bigg| \, , \qquad \mu = | \mu_t \mu_r | \,. 
\end{equation}
For the SIS lens, the infinitesimal axial ratio and magnification are the same, as the radial eigenvalue is unity. From Eqs.~(\ref{eq:lambdas}) and (\ref{lambdasis}) they are given by
\begin{equation}
\label{eq:mag}
R_\lambda =\mu = \left\{ \begin{array}{lc}
\dfrac{1}{x} - 1, \  \ \text{if $x\leq 1$} \\
\\
1-\dfrac{1}{x}, \  \ \text{if $x>1$}. \\ 
\end{array} \right.
\end{equation}
Using the lens equation (\ref{eq:sislenseq}), these quantities are expressed in terms of the source plane variables as
\begin{equation}
R^{\rm ex,in}_{\lambda} =\mu^{\rm ex,in} = \pm 1 + \frac{1}{s},
\label{mus}
\end{equation}
were $s = y$ is the position of the source and the positive sign corresponds to the external image and the minus sign to the internal one.

\subsection{Analytic solutions for arcs from circular sources}
\label{sec:imageformation}

\begin{figure*}
\includegraphics[width = \textwidth]{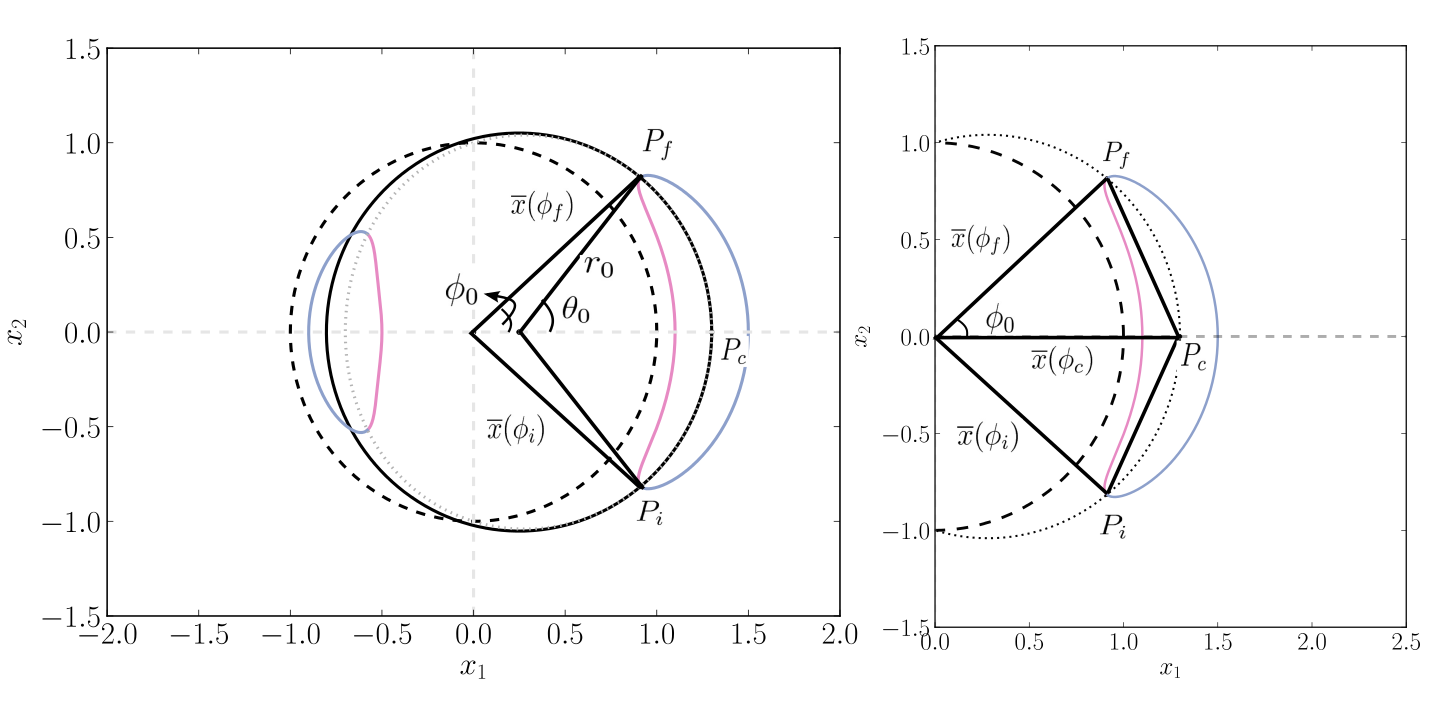}
\vspace*{-0.5cm}
\caption{Sketch showing the inner and outer parts of the external and internal arcs (magenta and blue curves), arc rigdeline (dotted line on the right panel) and critical curve (dashed line). The points 
$P_i$, $P_f$ and $P_c$ are 
the extremities and center of the arc, whose radial position 
is the ridgeline evaluated at the angular positions for these points ($\overline{x}\left(\phi_i\right)$, $\overline{x}\left(\phi_f\right)$ and $\overline{x}\left(\phi_c\right)$).
On the left panel we also show the circumference of radius $r_0$ passing through these 3 points, 
which defines the curvature center,
and the arc aperture $2\theta_0$ with respect to this center 
(black solid line).}
    \label{fig:sis_diagramaL}
\end{figure*}

We use the lens equation to map a set of points representing the contour of a 
circular source so as to obtain its images. 
Consider 
a circular source with radius $R_0$ centered at $\left(s_1, s_2 \right)$. We may write its boundary in the source plane as 
\begin{equation}
\label{eq:r0}
	\mathbf{R_0} = \mathbf{y} - \mathbf{s},\  \   \text{with}\  \   \mathbf{s} = s_1 \mathbf{\hat{y}_1} + s_2 \mathbf{\hat{y}_2}.
\end{equation}
Substituting the lens equation~(\ref{eq:sislenseq}) in the expression above and changing to polar coordinates yields
\begin{equation}
	\mathbf{R_0} = \left(x-1-s_{1}\cos\phi -s_{2}\sin\phi\right)\mathbf{\hat{x}} +\left(s_{1}\sin\phi - s_{2}\cos\phi\right)\mathbf{\hat{\phi}},
\end{equation}
where $\phi$ is the polar angle.
Using $R_0^2 = \vert \mathbf{R_0}\vert ^2$ 
and solving for $x$, we obtain
\begin{equation}
	x^{(\pm)} = 1 + s_{1}\cos\phi + s_{2}\sin\phi \pm \sqrt{R_0^2-\left(s_{1}\sin\phi - 		    s_{2}\cos\phi\right)^2}.
	\label{eq:xmaismenos}
\end{equation}
Rewriting $s_1$ and $s_2$ as 
\begin{equation}
s_1 = s\cos \theta, \hspace*{1cm} s_2 = s\sin\theta,
\end{equation}
Eq.~(\ref{eq:xmaismenos}) becomes
\begin{equation}
x^{(\pm)} = 1 + s\cos(\phi-\theta) \pm \sqrt{R_0^2 - s^2\sin^2(\phi-\theta)},
\label{eq:xmaismenosstheta}
\end{equation}
which gives the outer and inner parts of the images, as indicated  in Fig.~\ref{fig:sis_diagramaL}. 

The points where the discriminant in Eq.~(\ref{eq:xmaismenosstheta}) is zero define the arc extremities. Two ranges of $\phi$ may have a positive discriminant, indicating the existence of two solutions 
(as in the example of Fig.\ref{fig:sis_diagramaL}),
one inside the tangential critical curve and the other outside it, which we refer to as the 
internal and external arcs, representing them with the upper labels ``in'' and ``ex'' along the paper.

Eq.~(\ref{eq:xmaismenosstheta}) is the well known analytic solution for circular sources and the SIS lens \citep[see e.g.][and D\'umet-Montoya et al., in prep., for more generic solutions in isothermal models]{2005ApJ...634...77I,Dobler2006,Dobler2008}.
An expression of similar form is obtained as an approximate solution for arcs for generic radial profiles in the perturbative method of \citet{Alard2007}, which is exact in the SIS case \citep{Habib2013}. Therefore, we expect that the approach of this paper can be extended for more generic lens models, 
either using exact or perturbative solutions.

We define the arc ridgeline\footnote{
Rigorously speaking the arcs we are considering do not have a ridgeline, as they represent only a boundary (or the image of a uniform brightness source).
However, for a source with radial brightness distribution (i.e. with concentric circular isophotes) the brightness peak along any radial direction will be given by the curve defined in Eq.~(\ref{eq:ridgeline}). Therefore we employ this nomenclature even in the current case.}
as the mean of the inner and outer parts of the arc, which is independent of the source radius and is given by
\begin{equation}
\overline{x}(\phi) = \dfrac{x^{(+)}+x^{(-)}}{2} = 1+ s\cos(\phi-\theta).
\label{eq:ridgeline}
\end{equation} 
This curve is also shown in Fig.~\ref{fig:sis_diagramaL} (dotted line), which contains the tangential critical curve as well (dashed circle).

The curve given by expression~(\ref{eq:ridgeline}) is known as the Pascal lima\c{c}on \citep{limacon}. If $s \leq 1/2$, the lima\c{c}on is convex; if $1/2<s<1$, the lima\c{c}on is dimpled; if $s=1$, the lima\c{c}on degenerates to a cardioid and if $s>1$, the lima\c{c}on has an inner loop. The lima\c{c}on is not a circumference, therefore the arc ridgeline is not an arc segment. On the other hand, the portion passing across the external arc is quite similar to an arc of a circle, but with curvature center shifted from the lens' center (see Sec. \ref{sec:GeometricalProperties} and the Appendix \ref{sec:arcellipse}).

Since the lens is axially symmetric we can choose the source position along the (positive) x-axis, such that we set $\theta = 0$ without loss of generality.
In this case the center of the external and internal arcs will be at $\phi_c = 0$ and $\pi$, respectively. 
From the discriminant in Eq.~(\ref{eq:xmaismenosstheta}) we obtain the angular position for the arcs extremities, 
which, for the external arc, are given by
$\phi_i = - \phi_0$ and 
$\phi_f = \phi_0$, where 
\begin{equation}
\phi_0 = \arcsin \left(\frac{R_0}{s}\right).
\label{eq:phi0}
\end{equation}
For the internal arc we simply add $\pi$ to both angles, since they are complementary. 

In this paper we are interested in the case of images that can have large magnifications and length-to-width ratios, which implies that the sources must be smaller than the Einstein radius, $R_0 < 1$.
There are three possible image configurations for the finite sources in this case: Einstein ring, two images or one image. 
These configurations depend on the position $s$ of the center of the source relative to its radius $R_0$:
\begin{itemize}
\item  for $s \leq R_0 $ the source includes the tangential caustic, i.e. the lens center, and we have an Einstein ring. If $s= 0$ the ring is centered at the origin. The equality indicates the limit between the formation of two images and an Einstein ring,
where the two images touch at their extremities.
\item for $R_0< s < 1+R_0$ the source is inside the pseudo-caustic and in this case we have two images.
\item for $s \geq 1+R_0$ the source is completely outside the pseudo-caustic and in this case we have one image. 
\end{itemize}

The external image is always arc-shaped, until the formation of the Einstein ring. The situation with the internal image is a bit more tricky, as its shape will depend on the values of $s$ and $R_0$. The arc rigdeline (i.e. the lima\c{c}on) can provide a guideline to classify this image.
For $s<1/2$, i.e. when the source center position is smaller than half the pseudo-caustic radius, the lima\c{c}on is smooth and has positive curvature in all positions. 
In this case, the image will have an arc shape, as in Fig.~\ref{fig:sis_diagramaL} (left panel). 
At $s=1/2$ the lima\c{c}on has zero curvature at the image position and the arc ridgeline will be straight for small sources. 
In the intermediate region, $1/2<s\leq 1$, where the lima\c{c}on has a cusp towards the center, this image will lose its arc shape and starts looking like a ``droplet'' (round on one side and more pointy towards the center of the lens), becoming larger in the radial than in the tangential direction. 
For $s>1$ only the outer solution of the internal arc will be real, as this image will contain the lens center. 
This second image would not exist in the case of an infinitesimal source, but is present for a finite source, provided that the previously discussed condition, 
$R_0< s < 1+R_0$,
holds. 

Understanding 
the shape
of the internal image is important for interpreting the results of Section~\ref{sec:GeometricalProperties}. In particular, some length and width definitions will start to have an odd behavior for $s \gtrsim 1/2$.
The magnification of the internal image is close to unity at $s = 1/2$ and this image becomes highly demagnified as $s$ approaches $1$. 
In the present study we shall focus only on high magnifications and distortions, which occur for $s < 1/2$. However, we will still show some results for larger values of $s$ for completeness.

Both the external arc and the internal arc in the $s < 1/2$ regime have smooth extremities, as in the left panel of 
Fig.~\ref{fig:sis_diagramaL}. However, on the verge or merging and forming an Einstein ring, the extremities become sharp \citep[see e.g., Fig. 2 of][which considered finite circular sources and a point lens]{1964PhRv..133..835L}.

One may define the center of curvature of the arc(s) as the center of the circumference that passes through the arc extremities, $P_i$ and $P_f$, and its center, $P_c$ (see Fig.~\ref{fig:sis_diagramaL}). The position of this center is given by%
\footnote{From here on we will use the convention that the first sign (i.e. the $+$ in $\pm$ and the $-$ in $\mp$) will refer to the external arc, while the second one 
will correspond  to the internal image. The superscripts ``ex'' and ``in'' will only be kept whenever needed for clarity.}
\begin{equation}
x_{0} = \dfrac{s}{2}\left(1+\dfrac{1}{1\pm s\left( 1+ \cos \phi_0 \right)}\right),
\label{centeroffsets}
\end{equation}
such that the center of curvature of the arc(s) is offset with respect to the lens center. 
For $s \ll 1$ this offset is simply given by $s$, i.e., the center of curvature is at the source center.

The arc radius of curvature will be given by
\begin{equation}
\label{newr0}
r_{0} = \overline{x}(0) \mp x_{0} = 1 \pm \left(s-x_0\right).
\end{equation}

\section{Geometrical properties of the images}
\label{sec:GeometricalProperties}

In order to compute the magnification and arc cross sections, we need to determine the area and the axial ratio of the images. The next sections are devoted to the computation of these geometrical quantities and the search for accurate expressions in closed form.


\subsection{Area and finite source magnification}
\label{sec:Area}

The area of the image(s) can be written as:
\begin{equation}
\mathcal{A} = \int^{\phi_{f}}_{\phi_{i}}\int^{x^{(+)}}_{x^{(-)}} x \ dx\ d\phi 
=\int^{\phi_{f}}_{\phi_{i}} W(\phi) \overline{x}(\phi) \ d\phi,
\label{area}
\end{equation}
where $\overline{x}(\phi)$ is given in Eq.~(\ref{eq:ridgeline}) and 
\begin{equation}
W(\phi) = x^{(+)}-x^{(-)} = 2R_0\sqrt{1-\left(s/R_0\right)^2 \sin^2\phi},
\label{eq:wphi}
\end{equation}
i.e., $W(\phi)$ is the arc width measured in the radial direction at the angular position $\phi$. This quantity is the same for the internal and external arcs, as the SIS has no radial magnification.

For $s > R_0$ there are two ranges of $\phi$ (spanning $2\phi_0$ each), corresponding to the two images, and
the area for the external and internal arcs is given by:
\begin{equation}
\mathcal{A}= 4 R_0 \mathcal{E}\left(\phi_0, \left(s/R_0\right)^2\right) \pm \pi R_0^2,
\label{eq:Aex}
\end{equation}
where $\phi_0$ is given in Eq.~(\ref{eq:phi0}) and 
$\mathcal{E}(\alpha, m)$ is the incomplete elliptic integral of the second kind \citep{byrdhandbook}, which is given by 
\begin{equation}
\mathcal{E} (\alpha, m) = \int_0^\alpha \sqrt{1-m\sin^2 \theta}d\theta = \int_0^{\sin\alpha}\sqrt{\dfrac{1-mt^2}{\left(1-t^2\right)}}dt.
\end{equation}

The magnification of each image is simply the ratio of the image and source areas 
\begin{equation}
\label{eq:areaexact}
\mu_{\mathcal{A}}^{\rm ex,in} = \frac{\mathcal{A}}{\pi R_0^2} = \pm 1 + \frac{4}{\pi R_0} \mathcal{E}\left(\phi_0, \left(\frac{s}{R_0}\right)^2\right).
\end{equation}
This result is equivalent to the one found by \citet[][their Eq. (5)]{2005ApJ...634...77I}, which was expressed in terms of complete elliptical integrals of the first and second kinds.

The finite source magnification (\ref{eq:areaexact}) can be expanded for low values of $R_0/s$, i.e., far from the Einstein ring formation, as 
\begin{equation}
\label{eq:areaaprox}
\mu_{\mathcal{A}}^{\rm ex,in} \approx \pm 1 + \frac{1}{s} + \frac{R_0^2}{8s^3} + \mathcal{O}\left[\left(\frac{R_0}{s}\right)^4\right],
\end{equation}
which is valid even for very large magnifications ($s \ll 1$).
As expected, at zeroth order in the source size, the finite source 
magnification is exactly the Jacobian of the transformation, 
$\mu$ (Eq. \ref{mus}). 
The first correction for finite size is quadratic in $R_0$.
We notice that, for small sources and before the formation of an Einstein ring,
the effect of finite source is always to increase the magnification with respect to the infinitesimal case, for both the internal and external arcs. 
This same qualitative result was found in \citet{1979ApJ...233..402B} for a point mass lens.

The total magnification of the source is the ratio of the area of all images to the source area, which is simply the sum of the magnifications for each arc 
\begin{equation}
\label{eq:totalmag}
\mu_{\mathcal{A}} = \frac{\mathcal{A}^{\rm in} + \mathcal{A}^{\rm ex}}{\mathcal{A}_s} = \frac{8 }{\pi R_0} \mathcal{E}\left(\phi_0, \left(\frac{s}{R_0}\right)^2\right). 
\end{equation}

In this work we are interested in highly magnified (and distorted) sources. For the SIS this happens only in the regime where there are two images or an Einstein ring and for sources smaller than the Einstein radius ($R_0<1$).\footnote{The magnification of large sources ($R_0>1$) is discussed in \citet{2005ApJ...634...77I}.} Therefore, we will use expression (\ref{eq:totalmag}) throughout the paper.

In the case of Einstein rings ($s \leq R_0$) the integral in (\ref{area}) runs from $0$ to $2 \pi$, so that
$\phi_0 = \pi/2$ in the expression above, and the area is
\begin{equation}
\mathcal{A}^{{\rm E}}= 
8 R_0 \, \mathcal{E}\left(\left(\frac{s}{R_0}\right)^2\right),
\label{eq:Aring}
\end{equation}
where $\mathcal{E}(m)=\mathcal{E}(\pi/2, m)$ is the complete elliptic integral of the second kind. 
Therefore the Einstein ring magnification is
\begin{equation}
\label{eq:Eringmag}
\mu_{\mathcal{A}}^E=\frac{8}{\pi R_0} \mathcal{E}\left(\left(\frac{s}{R_0}\right)^2\right), 
\end{equation}
which is again equivalent to the result in  \citet[][]{2005ApJ...634...77I}.

Close to a centered
Einstein ring ($s/R_0 \ll 1$) the magnification can be expanded as
\begin{equation}
\label{eq:Eringmagapprox}
\mu_{\mathcal{A}}^E \approx \frac{4}{R_0}- \frac{s^2}{R_0^3} + 
\mathcal{O}\left[\left(\frac{s}{R_0}\right)^4\right].
\end{equation}
The first term corresponds to the magnification of a perfectly aligned observer--lens--source and gives the maximum magnification for a finite circular source, $\mu_{\max} = 4/R_0$ \citep{1982MNRAS.199..987P}. In this case the image is an annulus of circumference $2 \pi$ and width $2 R_0$.

Combining Eqs.~(\ref{eq:totalmag}) and (\ref{eq:Eringmag}), for $s/R_0>  1$ and $\le 1$, respectively, gives the total magnification for the whole range of source positions $s$. 
The maximum value of this function occurs for $s=0$ and is given by $\mu_{\max}$. In the boundary between the two arc and the Einstein ring solutions the magnification is $\mu_{\rm trans} = 8/(\pi R_0)$.


\subsection{Length}
\label{sec:Length}

Contrarily to the area, there is no unique definition of length for generic shapes. In the case of gravitational arcs a few choices have been used in the literature for both simulated and real images. They all use the arc extremities, points $P_i$ and $P_f$ in Fig.~\ref{fig:sis_diagramaL}, and involve the determination of an arc center in a way or another. This center is usually chosen as the image of the center of the source, in the case of simulated images, corresponding to $P_c$ in this figure. 
Below we test several length definitions, seeking at the same time expressions that are accurate to describe the arc shape and that are written in a simple form in terms of elementary functions. 

\begin{itemize}
\item[1)] Geometrically, the simplest length definition is to consider the sum of the segments connecting the arc extremities to its center (see Fig.~\ref{fig:sis_diagramaL} right panel):
\begin{equation}
 L_{1} = \overline{P_iP_c}+ \overline{P_cP_f}. 
\end{equation}
This definition has been applied to both real and simulated arcs
\citep[see e.g.,][]{2002ApJ...573...51O,2005APh....24..257H,Xu2016}. The angular positions of
$P_i$, $P_f$ and $P_c$ are simply $\phi_i=-\phi_0$, $\phi_f=\phi_0$ and $\phi_c =0$, and their radial position is simply the ridgeline 
evaluated at these angles, $\overline{x}\left(\phi_i\right)$, $\overline{x}\left(\phi_f\right)$ and $\overline{x}\left(\phi_c\right)$. Therefore, the lengths of the external and internal arcs will be 
given by
\begin{equation}
L_{1} = 2 \sqrt{2+R_0^2 - 2\cos \phi_0 \pm 2R_0 \sin \phi_0}.
\label{L1}
\end{equation}

\item[2)] A simple way to define a length that follows the shape of the image is to integrate the tangential part of the ridgeline (Eq. \ref{eq:ridgeline}) along the arc: 
\begin{equation}
L_2 = \int ^{\phi_f}_{\phi_i} \overline{x}(\phi) d\phi = 2\left(\phi_0 \pm R_0\right) = 
2\left(\arcsin \left(\frac{R_0}{s}\right)\pm R_0\right),
\label{L2}
\end{equation}
which has also a very simple expression.

\item[3)] More rigorously, the length of the arc ridgeline is given by
\begin{equation}
L_3 = \int dl = \int_{\phi_i}^{\phi_f} \sqrt{\overline{x}^2 + \left(\dfrac{d \overline{x}}{d\phi}\right)^2}d\phi.
\label{L3}
\end{equation}
This is the most natural definition of an ``exact'' arc length in this context and will be taken as a reference when we compare the different expressions for the length that will be tested. As far as we know, the only application of this definition to arcs is given by the so-called Mediatrix method \citep[][and Bom et al., in prep.]{2012arXiv1212.1799B,2017A&A...597A.135B}.

The expression above is the length of the lima\c con between the image extremities and is given by
 \begin{equation}
 L_3^{\rm ex} = 4\left(1+s\right)\mathcal{E}\left(\frac{\phi_0}{2},\dfrac{4 s}{\left(1+s\right)^2}\right)
 \end{equation}
 and
\begin{equation}
L_{3}^{\rm in} = 4\left(1+s\right)\left[\mathcal{E}\left(\dfrac{\pi+\phi_0}{2},\dfrac{4s}{(1+s)^2}\right)- \mathcal{E}\left(\dfrac{4s}{(1+s)^2}\right)\right].
\end{equation}
Although several numerical methods exist for the fast computation and inversion of these functions \citep[see e.g.,][]{FUKUSHIMA201343, FUKUSHIMA201571}, the expressions above do not allow us to obtain the arc cross section in a simple form. Therefore, we will seek other definitions that provide results close to the one above, but can be expressed in terms of simple functions. 

\item[4)] Currently, the most commonly used length definition \citep[see e.g.,][]{1993ApJ...403..509M,Bartelmann,Meneghetti2008} is given by the arc of circumference passing through the image extremities and its center (i.e. the circle containing the points $P_i$, $P_c$ and $P_f$):
\begin{equation} \label{l4_def}
L_4 = 2\theta_0r_0,
\end{equation}
where $r_0$ is the curvature radius given by Eqs.~(\ref{centeroffsets}) and (\ref{newr0}) and $\theta_0$ is half of the arc aperture with respect to the curvature center, as indicated in Fig.~\ref{fig:sis_diagramaL}, and is given by
\begin{equation}
\label{theta0}
\theta_{0} = \arcsin\left(\dfrac{\left(1\pm \sqrt{s^2-R_0^2}\right)}{r_0}\frac{R_0}{s}\right).
\end{equation}

As we shall see, $L_4$ provides an excellent approximation to $L_3$ and is written explicitly in terms of simple functions. Nevertheless, it does not allow one to obtain the cross section in closed form. Therefore, we test two alternative length definitions using the arc of a circle, but now centered at the origin (lens center) instead of the curvature center, such that the arc spans the angle $2\,\phi_0$. The points $P_i/P_f$ and $P_c$ are located at different radii with respect to that center and we test with these two radii, defining $L_5$ and $L_6$.

\item[5)] Arc of a circle with aperture $2\,\phi_0$ and radius at $P_c$:
\begin{equation}
L_{5} = 2\phi_0\overline{x}\left(0\right)= 2\phi_0\left(1\pm s\right) = 2 \arcsin \left(\frac{R_0}{s}\right) \left(1\pm s\right), 
\end{equation}
which has indeed a simpler expression than $L_4$.

\item[6)] Same as above, but using as radius the distance between lens center and the arc extremities:
\begin{align}
L_{6} &= 2\phi_0 \overline{x}\left(\phi_{i}\right) 
= 2\phi_0 \left(1 \pm s\cos\phi_0\right)  \\
&=  2 \arcsin \left(\frac{R_0}{s}\right) \left(1 \pm \sqrt{s^2-R_0^2}\right),
\label{L6}
\end{align}
\end{itemize}
which is also more tractable than $L_4$.

\begin{figure*}
	\includegraphics[width=\textwidth]{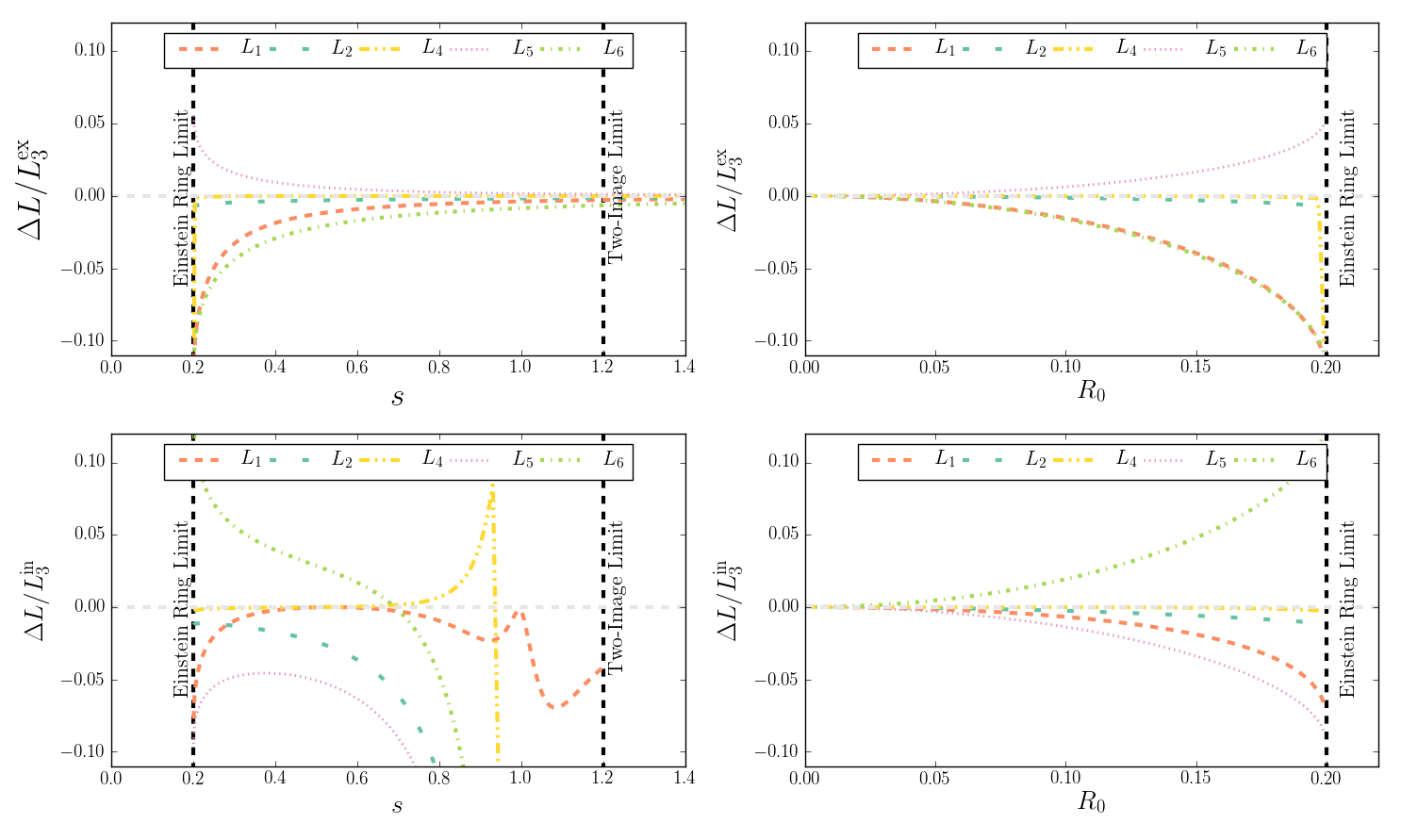}
    \caption{Relative difference ($\Delta L/L_3 = (L_i - L_3)/L_3$) between the various length measurements considered. The parameters used were: $R_0 = 0.2$ (left) and $s=0.2$ (right).}
    \label{fig:sis_length}
\end{figure*}

In Fig.~\ref{fig:sis_length} we show the fractional difference of all length definitions above as compared to $L_3$, which we take as the ``exact'' length. Let us focus first on the external arc (upper panel). We see that all proposed definitions are in excellent agreement in the whole range of $s$ and $R_0$. The highest deviations occur close to the Einstein ring limit and are at most of the order of $10\%$ for $R_0 = 0.2$.
In the case of $L_1$, as it approximates the arc by two chords, even for infinitesimal sources we have $L_1 \rightarrow 2\sqrt{2}$ close to the formation of the Einstein ring, such that the fractional difference with respect to $L_3$ is 
$\sim 10\%$ at this point. 
The approximations become better for smaller arcs (higher $s$, lower $R_0$), as will be discussed below. We see that $L_1$ and $L_6$ always underestimate the arc length, while $L_5$ overestimates it, which is the expected behavior from their definitions.
The expressions $L_2$ and $L_4$ are in striking agreement with $L_3$ all the way down to very close to the Einstein ring formation.

For the internal image the situation is a bit more complicated, since its shape can deviate substantially from an arc, which
happens somewhere in the interval $1/2<s<1$, depending on the source size, as discussed in Section~\ref{sec:imageformation}. 
The comparison of the length definitions in this case is shown in  Fig.~\ref{fig:sis_length} (lower panel). The changes in shape are clear from the behaviors as a function of $s$ in the bottom left panel. We recall that all length definitions are based on points along the lima\c{c}on between the extremities of the arc, which loses its concavity and becomes dimpled for $s>1/2$. For $s<1/2$ all length definitions are well behaved and similar, with less than $5\%$ deviation, except close to the Einstein ring formation. In the intermediate region the deviations become larger. For $s>1$ the length starts to decrease and some definitions cease to be valid.

To compute the arc cross section, we are interested in highly elongated and magnified images. The infinitesimal source $L/W$ (Eq.~\ref{mus}), which gives an order of magnitude of the finite source value, 
is unity at $s = 1/2$ and decreases for higher values of $s$. Therefore, we are only interested in the regime where the lengths 
are well defined and well behaved.
In this regime, all approximations agree to within $\sim 10\%$.
Again, the best approximations are $L_2$ and $L_4$, specially for lower values of $s$, which is the relevant regime for the cross section. 

In the lower right panel of Fig.~\ref{fig:sis_length} we show the lengths as a function of $R_0$ for $s = 0.2$. The behavior is qualitatively similar to that of the external arc, but now $L_5$ underestimates and $L_6$ overestimates the length, also as expected from their definitions.
It is clear that, by far, $L_2$ and $L_4$ are the best approximations to $L_3$ in all the relevant range of $s$ for both the external and internal arcs. In particular, $L_4$ is almost indistinguishable from our reference length definition. 
Nevertheless, we will use $L_2$ to compute the arc cross section in Sec. \ref{sec:sigmaLW} owing to its simplicity, which will allow us to obtain an expression in closed form.

The expressions (\ref{L1}--\ref{L6}) that we have obtained above for $L_1$--$L_6$ are valid for any source size $R_0$ and position $s>R_0$. However, it is useful to obtain perturbative solutions for small sources.
Expanding these expressions 
up to third order in $R_0$ we obtain:%
\footnote{Considering $s<1/2$ for the internal arc, for the reasons discussed previously on the behavior of the internal image around this region.} 
\begin{equation}
L_{1} \approx 2\left(\pm 1+\frac{1}{s} \right)R_0 + \frac{R_0^3}{4s^3(1 \pm s)},
\label{L1P}
\end{equation}
\begin{equation}
L_{2} \approx 2\left(\pm 1+\frac{1}{s} \right)R_0 +  \frac{R_0^3}{3s^3},
\end{equation}
\begin{equation}
L_{3} = L_4 \approx 2\left(\pm 1+\frac{1}{s} \right)R_0 + \frac{(1\pm s + s^2)R_0^3}{3s^3(1\pm s)},
\label{L3P}
\end{equation}
\begin{equation}
L_{5} \approx 2\left(\pm 1+\frac{1}{s} \right)R_0 + \frac{(1 \pm s)R_0^3}{3s^3} ,
\label{L5P}
\end{equation}
\begin{equation}
L_{6} \approx 2\left(\pm 1+\frac{1}{s} \right)R_0 +  \frac{(1\mp 2s)R_0^3}{3s^3}.
\label{L6P}
\end{equation}
These approximations are valid for arbitrary magnifications, as long as we are far from the Einstein ring formation ($R_0\ll~s$).
The first order term in $R_0$ yields exactly the infinitesimal source approximation $L = 2R_0\vert \mu_t\vert _{x = 1 \pm s}$ as expected (see Eqs. \ref{lambdasis} and \ref{mus}). The expansions above provide the lowest order corrections for finite sources to the various length definitions.

For small values of $s$, i. e. high magnifications, all proposed measures of $L$ (except $L_1$) also agree up to third order in $R_0$. Interestingly the expansions  
for $L_3$ and $L_4$ agree exactly up to this order, for any value of $s$, which is in agreement with what we see in Fig.~\ref{fig:sis_length}.
Neglecting the quadratic term in $s$ in expression (\ref{L3P}), we see that $L_2 = L_3 (L_4)$, which explains why $L_2$ is so close to $L_3 (L_4)$ in the plots of Fig.~\ref{fig:sis_length}.
It is also clear from the expressions (\ref{L3P}), (\ref{L5P}), and (\ref{L6P}) why $L_5$ is larger and $L_6$ is smaller than $L_3$ for the external arc, and the other way around for the internal one.
In brief, all the qualitative behaviors pointed out in Fig.~\ref{fig:sis_length}
are clearly seen in the perturbative expansions above.


\subsection{Width}
\label{sec:Width}
If the arc length has not a unique definition, the determination of the width is even more ambiguous. Several methods have been proposed and tested in the literature \citep[see e.g.,][for  reviews]{2012A&A...547A..66R,2013SSRv..177...31M}. As in the previous section, our aim here is to test several definitions of $W$ seeking  expressions that are at the same time representative of the arc shape and that can be expressed in simple analytical form.

A natural definition in the context of the smooth SIS arcs with a well defined boundary is to choose the width along the direction perpendicular to the ridgeline at the arc center:
\begin{equation}
W_{c} =W(0) = 2R_0,
\label{Wc}
\end{equation}
where $W(\phi)$ is given in Eq.~(\ref{eq:wphi}). The result above, which is the same for internal and external arcs, is easy to interpret as the lensing by a SIS does not change the radial positions and thus the width of the image defined as above is the same as the source diameter.

For more realistic arcs, from ray-tracing simulations or real data, the shapes can be less symmetrical and the object boundary is subject to irregularities. It is therefore suitable to use information from the whole object, instead of a measurement across a single direction, as above.
One approach that has been often used in the literature is to derive a width from the object area $\mathcal{A}$ and length $L$, $W \propto \mathcal{A}/L$. The proportionality constant depends on the shape of the object. In 
\citet{Bartelmann} and subsequent works, the images are fitted by simple geometric figures, such as rectangles, ellipses or rings. The figure that best fits the objects defines the constant, which is, for example, $1$ for rectangles and $4/\pi$ for ellipses. It turns out that the arcs that we consider in this paper are very well fit by a figure known as ArcEllipse \citep[][see Appendix \ref{sec:arcellipse}]{Furlanetto}. The $\mathcal{A}$--$L$--$W$ relation for the ArcEllipse is identical to that of an ellipse and therefore we define the width as
\begin{equation}
W_i = \dfrac{4\mathcal{A}}{\pi L_i},
\end{equation}
where $L_i$ represents the definitions of lengths used previously. As in the previous section, we take the length along the ridgeline $L_3$ and define $W_3$ as our reference value to compare the different width definitions.

Another width definition that has been used more recently \citep{Meneghetti2008,2012A&A...547A..66R} is to consider the mean (or the median) of the width of the object along the radial direction with respect to its center (or in the direction orthogonal to the object ridgeline). This is akin to computing
\begin{equation}
\overline{W} = \dfrac{1}{\phi_f - \phi_i}\int_{\phi_i}^{\phi_f} W(\phi)d\phi = \dfrac{2R_0}{\phi_0}\mathcal{E}\left(\phi_0, \dfrac{s^2}{R_0^2}\right).
\label{meanW}
\end{equation}

\begin{figure*}
	\includegraphics[width=\textwidth]{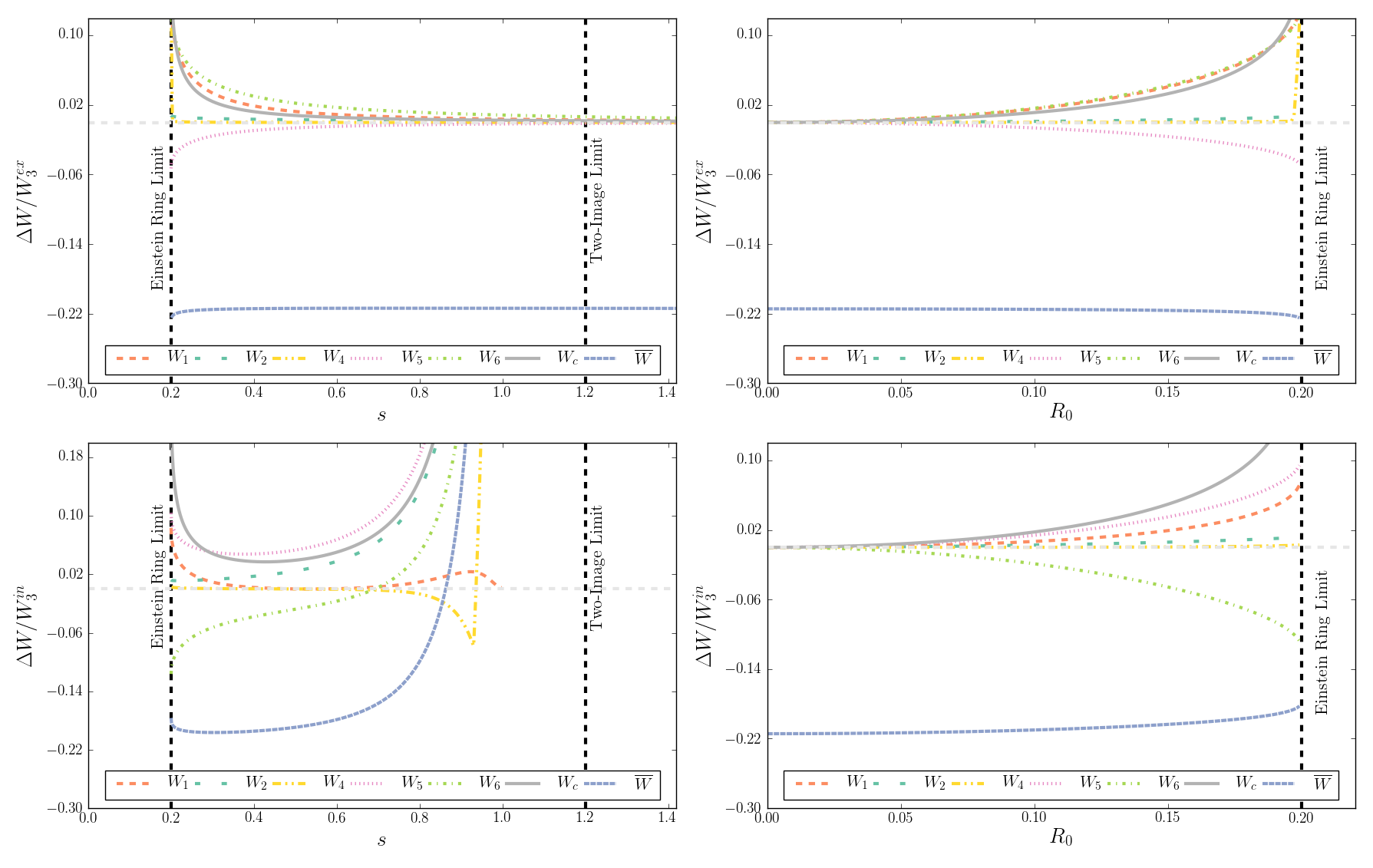}
    \caption{Relative difference ($\Delta W/W_3 = (W_i - W_3)/W_3$) between the various width measurements considered. The parameters used were: $R_0 = 0.2$ (left) and $s=0.2$ (right).}
    \label{fig:sis_width}
\end{figure*}

In
Fig.~\ref{fig:sis_width} we show the relative difference of the 
various definitions tested with respect to the reference value, $W_3$, as a function of $s$ and $R_0$.
All expressions, except $\overline{W}$, agree reasonably well in the whole interval of $s$ and $R_0$ (we recall that only the range $s<1/2$ is relevant for the internal arc).
The behavior of $W_i$ traces back to the behavior of $L_i$ seen in Section~\ref{sec:Length}, as $\mathcal{A}$ is the same for all width definitions.
We see that the ratio $\overline{W}/W_3$ is almost constant in the whole range of parameters (considering $s <1/2)$. The proportionality factor is discussed below and in Appendix~\ref{sec:arcellipse}. The expression $W_c$ is closer to $W_3$ for the external than for the internal one. The difference is at most $\sim 10\%$ ($\sim 20\%$), close to the Einstein ring limit, for the external (internal) arc, and less then $\sim 2\%$ ($\sim 5\%$) in most of the parameter range.
Owing to the simplicity of $W_c$, we will use this expression
to obtain the arc cross section in an analytical form. On one hand, the differences pointed out above with respect to $W_3$ are much smaller than the finite source effect on the cross section.
On the other hand, in a practical application, as long the arcs are measured in the same way as used to compute the cross section, any definition of $W$ is valid.

As we did for the length, it is illustrative to derive the perturbative expansions of the width for small source sizes, which are given by
\begin{equation}
W_1 \approx 2R_0 \pm \frac{R_0^3}{4s\left(1 \pm s\right)^2},
\end{equation}
\begin{equation}
W_2 \approx 2R_0 - \frac{R_0^3}{12s^2( 1 \pm s)},
\end{equation}
\begin{equation}
W_3 = W_4 \approx 2R_0 - \frac{(1 \pm s + 4s^2)R_0^3}{12s^2( 1 \pm s)^2},
\end{equation}
\begin{equation}
W_5 \approx 2R_0 - \frac{(1  \pm 4s)R_0^3}{12s^2(1 \pm s)},
\end{equation}
\begin{equation}
W_6 \approx 2R_0 - \frac{( 1  \mp 8s)R_0^3}{12s^2(1 \pm s)},
\end{equation}
\begin{equation}
W_c = 2R_0,
\end{equation}
\begin{equation}
\overline{W} \approx \frac{\pi R_0}{2} \left(1 - \frac{ (4 - 3s)R_0^2}{24s^2}\right).
\end{equation}
Here again the first term is derived from the eigenvalue of the Jacobian matrix of the transformation, $W = 2 R_0 \mu_r = 2 R_0$, and the first correction is quadratic in $R_0$ with respect to this term (except for $W_c$). The differences among the $W_i$ 
are due to the differences in $L_i$ and all $W_2$--$W_6$ agree for high magnifications ($ s \ll 1$) up to third order in $R_0$.
It is clear that $\overline{W}$, when corrected by a factor $4/\pi$, gives the same result as the other definitions 
(to first order in $R_0$).
The same happens for an ArcEllipse (Appendix~\ref{sec:arcellipse}), for which the relation $W_c/\overline{W}=4/\pi$ holds exactly. The same correction factor was found by \citet{2012A&A...547A..66R} to relate the mean width to the one based on the area of the ellipse.

To derive a perturbative expression for the arc cross section for small sources, it is useful to write the length-to-width ratio from the expansions that were obtained before. In particular, as we will compute the cross section based on $L_2$ and $W_c$, we obtain the ratio
\begin{equation}
\label{LWapprox}
\frac{L_2}{W_c}\approx \pm 1 + \frac{1}{s} + \frac{R_0^2}{6 s^3},
\end{equation}
which yields the infinitesimal axial ratio $R_\lambda$ (Eq.~\ref{eq:Rlambda}) at zeroth order in $R_0$ and the lowest order correction for finite sources.

As in the case of the magnification, the axial ratio is always  increased with respect to the infinitesimal source, as long as we are far from the Einstein ring formation ($s \gg R_0$).


\section{Cross Sections}
\label{sec:Cross Section}

The cross section is defined as the area in the source plane that generates images with some specified 
properties, e.g. axial ratio or magnification above a certain threshold \citep{Schneider,1995A&A...297....1B}:%
\footnote{
We recall that all distances in this paper are given in units of the Einstein radius. Therefore, to convert the cross section to physical units, one must multiply this expression by $\xi_0^2$ (from Eq.~\ref{RE}). In terms of the solid angle in steradians: 
$\sigma_{\rm sr} = \sigma \left(\xi_0/D_{\rm OL}\right)^2.$}
\begin{equation}
\label{eq:sigdef}
\sigma = \int_{\Omega} d^2y,
\end{equation}
where the domain $\Omega$ is the region in the source plane satisfying the condition, for example, magnification above a given threshold ($\mu > \mu_{\rm th}$) or images with length-to-width ratio above a given value ($L/W > R$).

The cross section can be expressed in terms of the lens plane variable $\mathbf{x}$ using the Jacobian (\ref{Jaco}) 
\begin{equation}
\label{eq:siglens}
\sigma = \int_{\Omega_x} \vert \det \mathbb{J}(x)\vert d^2x,
\end{equation}
where now the domain of integration is defined for the quantities (e.g. length-to-width ratio or magnification) expressed in terms of the lens plane coordinates. 

As the local magnification and axial ratio (Eqs.~\ref{eq:lambdas} and \ref{eq:Rlambda}) are naturally obtained in the lens plane, the form of the cross section above is the most often used when considering infinitesimal sources. Working in the source
plane is computationally more expensive, as it requires solving the lens equation (\ref{eq:yx}). Care must be taken
when working in the lens plane 
as multiple regions in this plane (corresponding to different images) can be mapped to the same region in the source plane. This multiplicity has to be accounted for in the cross section computation.

When finite sources are considered, the lens equation has to be solved (either numerically or analytically) to obtain the images, and the cross section is computed as in Eq. (\ref{eq:sigdef}).\footnote{See, however, \citet{Fedeli} and the discussion in Appendix \ref{sec:MagxR}.}
In the case of this paper, not only 
the lens equation has a simple analytical solution, but also we have derived expressions for $\mu$ and $L/W$ as a function of the position of the center of the source $s$ in closed form. This will enable us to compute the cross section (Eq. \ref{eq:sigdef}) in a simple form. 

In the SIS case with circular sources, due to the axial symmetry and since $\mu$ is a monotonically decreasing function of $s$, the cross section (\ref{eq:sigdef}) is simply given by 
\begin{equation}
\sigma_\mu = \pi s_{\rm th}^2,
\label{eq:sigSIS}
\end{equation}
where $\mu \left(s_{\rm th}\right) = \mu_{\rm th}$. 

For the arc cross section there is another condition, as for $s <R_0$ an Einstein ring is formed. Therefore, the domain in $s$ where arcs with $L/W>R$ are formed is given by $ s_{\rm th} \ge s > R_0$, where $L/W \left(s_{\rm th}\right) = R$. The cross section is thus the area of the annulus defined by this condition
\begin{equation}
\sigma_{L/W} = \pi s_{\rm th}^2 - \pi R_0^2.
\label{eq:sigSISannulus}
\end{equation}

\subsection{Infinitesimal Cross Section}

In the infinitesimal circular source approximation the axial ratio is given by Eq.~(\ref{eq:mag}), so that the condition $L/W = R_{\lambda} > R$ yields two solutions for $x$:
\begin{equation}
x_\lambda = \left\{ \begin{array}{lc}
x_{\max} =\dfrac{R}{R-1}, \\ 
\\
x_{\min} =\dfrac{R}{R+1}. \\ 
\end{array} \right.
\end{equation}
Therefore, the domain of integration in Eq.~(\ref{eq:siglens}) is an annulus with radii determined by the values above. Using the Jacobian given in Eq.~(\ref{eq:mag}), the cross section (\ref{eq:siglens}) is
\begin{eqnarray}\nonumber
\sigma_{\lambda} &=& \int_0^{2\pi} \int_{x_{\rm min}}^1 \left(\dfrac{1}{x}-1\right)x dx d\phi + \int_0^{2\pi} \int^{x_{\rm max}}_1 \left(1-\dfrac{1}{x}\right)x dx d\phi\\
&=& 
 \frac{\pi}{\left(R+1\right)^2} + \frac{\pi}{\left(R-1\right)^2}
= 2\pi \dfrac{R^2+1}{\left(R^2-1\right)^2},
\label{eq:sigmaRinf}
\end{eqnarray}
as obtained in 
\citet{1995A&A...297....1B,Habib2013,2013MNRAS.430.1423E}.
Notice that the region with $x<1$ corresponds to the internal image and the region with $x>1$ corresponds to the external image. Therefore, we can split the cross section into two, one for the internal image having $R_{\lambda} > R$ and the other for the external image satisfying this condition:
\begin{equation}
\sigma_\lambda^{\rm in,ex} = \frac{\pi}{\left(R \mp 1\right)^2}.
\label{eq:sigmainex}
\end{equation}
This result is the same one would have obtained working in the source plane Eq.~(\ref{eq:sigdef}) and considering the axial ratio for each image in the source plane (Eq.~\ref{mus}) to define the integration domain.

The total magnification is given, in the source plane (Eq.~\ref{mus}), simply by\footnote{This expression is valid for $s\leq 1$, so that there are two images, and therefore for $\mu \ge 2$.}
\begin{equation}
\mu = \mu^{\rm in} +  \mu^{\rm ex} = \frac{2}{s}.
\end{equation}
The condition $\mu > \mu_{\rm th}$ sets the cross section (Eq.~\ref{eq:sigSIS}) as \citep{Schneider}:
\begin{equation}
\sigma_{\mu} = \frac{4\pi}{\mu_{\rm th}^2}.
\label{sigmaginf}
\end{equation}

\subsection{Finite Source Magnification Cross Section}

\begin{figure*}
\includegraphics[width = \textwidth]{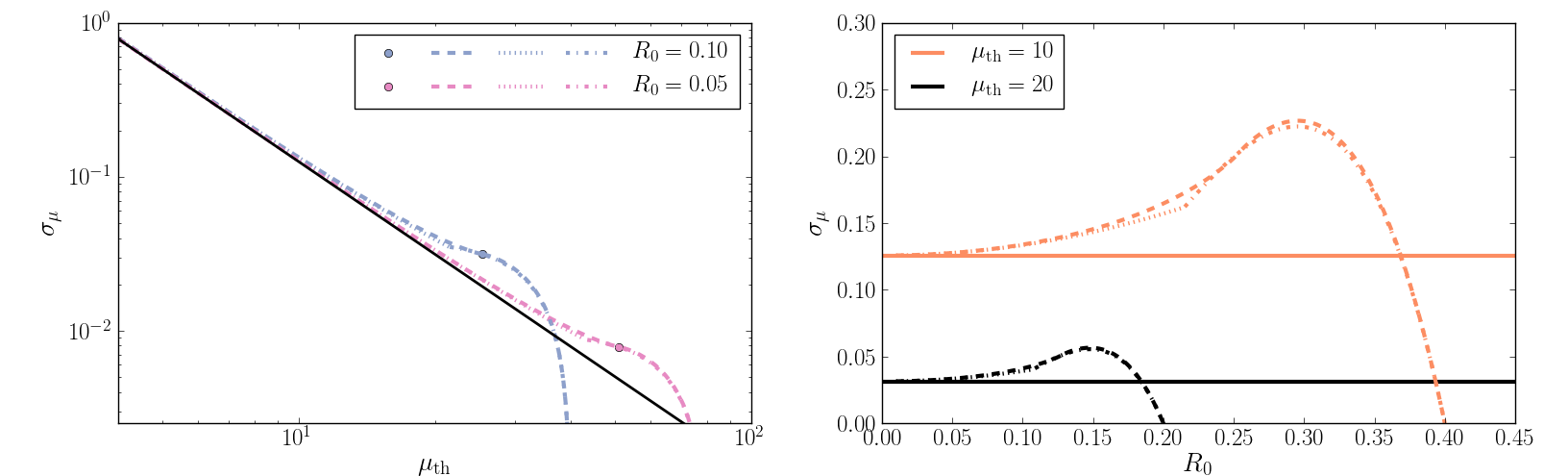}
\caption{Magnification cross section as a function of the magnification threshold (left) and source radius (right). The dashed lines represent the exact result from the numerical inversion of the elliptic integral. The small circles on the left panel correspond to the value of $\mu_{\rm trans}$, at which the solution switches from two arcs to an Einstein ring.
The dotted lines show the results of the perturbative expansions (\ref{sigmagsmall}) and (\ref{sigmaEring}). In the right panel, the 
larger dot on the dotted line corresponds to the value $\mu_J$ at which the approximate cross section transitions from the two perturbative expressions. The solid line is the infinitesimal source approximation (Eq. \ref{sigmaginf}).}

\label{fig:sigtot}
\end{figure*}

The exact magnification cross section for finite circular sources is found from (\ref{eq:sigSIS}) by solving $\mu_{\mathcal{A}} \left(s_{\rm th}\right) = \mu_{\rm th}$ for $s_{\rm th}$, where $\mu_{\mathcal{A}}$ is given by Eq. (\ref{eq:totalmag}) for $s\ge R_0$  and by Eq. (\ref{eq:Eringmag}) for $s\le R_0$.
The result from the numerical inversion of the elliptic integrals is shown in Fig.~\ref{fig:sigtot}, along with the results for infinitesimal sources (Eq. \ref{sigmaginf}) and an approximate solution discussed below.

For $R_0 \ll s$ 
we may obtain an analytical solution by using the approximation~(\ref{eq:areaaprox}), such that the total magnification is 
\begin{equation}
\mu_{\rm tot}^P = \frac{2}{s} + \frac{R_0^2}{4 s^3}. 
\label{magtotP}
\end{equation}
To determine $s_{\rm th}$ we solve the third order equation $\mu_{\rm tot}^P = \mu_{\rm th}$ and expand the solution to the lowest non-trivial order in $R_0$, to obtain the cross section 
\begin{equation}
\sigma_{\mu}^P = \frac{4\pi}{\mu_{\rm th}^2} + \frac{\pi}{4} R_0^2.
\label{sigmagsmall}
\end{equation}
As expected, in this regime the cross section for finite sources is enhanced with respect to the infinitesimal one, as the magnification is also higher in this case (Eq.~\ref{magtotP}).

We may also obtain a perturbative solution close to the perfectly aligned Einstein ring from Eq.~(\ref{eq:Eringmagapprox}), which can be easily solved for $s_{\rm th}$ to obtain
\begin{equation}
\sigma_\mu^{\rm E}= \pi \left(4 R_0^2 - \mu_{\rm th} R_0^3 \right) = \pi \left(\mu_{\max} - \mu_{\rm th} \right) R_0^3.
\label{sigmaEring1}
\end{equation}
The cross section vanishes for $\mu_{\rm th} \ge \mu_{\max}=4/R_0$, as no image can have a magnification above this value. This is in contrast to the infinitesimal source case, for which the magnification is unbounded and the cross section (Eq.~\ref{sigmaginf}) never vanishes. Thus, for high magnifications, within the Einstein ring regime, the finite source cross section is smaller than the infinitesimal one.

While the approximation (\ref{sigmagsmall}) is still good close to the onset of the Einstein ring formation (i.e. at $\mu = \mu_{\rm trans}$), the approximation (\ref{sigmaEring1}) breaks down at this point. However, it is easy to improve the Einstein ring cross section, considering that the expression above is linear in  $\left(\mu_{\max} - \mu_{\rm th} \right)$. We add a correction term that is quadratic in this quantity and fix the cross section at $\mu_{\rm trans}$ to its exact value ($\pi R_0^2$). In other words, we build an extreme perfect quadratic approximant to the cross section between the onset of the Einstein ring formation and the perfect Einstein ring solution, such that the cross section is
\begin{align}
\sigma_\mu^{\rm E}= \pi \left(4 R_0^2 - \mu_{\rm th} R_0^3 \right) - \frac{1}{R_0^2}\left(\frac{3 \pi -8}{(4-8/\pi)^2}\right) \left(4 R_0^2 - \mu_{\rm th} R_0^3\right)^2.
\label{sigmaEring}
\end{align}
In fact, this expression 
provides an excellent approximation for the magnification cross section in the full range from $\mu_{\rm trans}$ to $\mu_{\max}$, as can be seen in fig. \ref{fig:sigtot}.
By joining this solution with the expression (\ref{sigmagsmall}) we may construct a single continuous approximation to $\sigma_{\mu}$.  
These two curves match at $\mu_J = 2.15/R_0 < \mu_{\rm trans}$.

Therefore, we build a single approximate magnification cross section for finite sources in the full range of the magnification threshold 
by using $\sigma_{\mu}^P$ (Eq.~\ref{sigmagsmall}) for $ \mu_{\rm th} < \mu_J$ and $\sigma_{\mu}^E$ (Eq.~\ref{sigmaEring}) for $ \mu_{\rm th} \ge \mu_J$. This is shown as the dotted line curves in fig.~\ref{fig:sigtot}.

As we can see from Eqs. (\ref{eq:totalmag}) and (\ref{eq:Eringmag}), the magnification can be written as $\mu_{\mathcal{A}} = f\left(s/R_0\right)/R_0$, where $f$ is expressed in terms of the incomplete and complete elliptic integrals for $s\ge R_0$ and $s\le R_0$, respectively. Therefore, the cross section will be given by $\sigma_{\mu}= \pi \left[f^{-1}\left(\mu_{\rm th}R_0\right)\right]\ R_0^2$.
This form is explicit in Eqs.~(\ref{sigmagsmall}) and (\ref{sigmaEring})
and this is why $\mu_J$, $\mu_{\rm trans}$, $\mu_{\max}$,  etc. are all $\propto R_0^{-1}$. Given this form, the accuracy of approximations (\ref{sigmagsmall}) and (\ref{sigmaEring}), more specifically their fractional deviation with respect to the exact result, 
will be a function of the combination $\mu_{\rm th}R_0$. The highest discrepancy between the exact and approximate solutions 
occurs at $\mu_J$ 
and is $6.2\%$.
Outside the range $ 1.6 \lesssim \mu_{\rm th}R_0 \lesssim 2.3$ the approximations deviate less than 2\%.
In particular, this precision holds in the whole interval of magnifications within the Einstein ring formation. The perturbative solution is practically exact for $\mu_{\rm th}\lesssim R_0^{-1}$. Of course the approximate solutions can be improved arbitrarily by considering higher order expansions. However, the error achieved with expressions (\ref{sigmagsmall}) and (\ref{sigmaEring}) is already much smaller than other uncertainties involved in the modeling of the statistics of highly magnified sources
\citep[e.g.,][]{2011ApJ...734...52H,2012ApJ...755...46L}.

\subsection{Cross Section for Arc Formation}
\label{sec:sigmaLW}

In sections
\ref{sec:Length} and \ref{sec:Width} we have tested several definitions and approximations for $L$ and $W$, seeking  expressions that are at the same time accurate and written in a simple analytic form. In particular, we have found that $L_2$ (Eq.~\ref{L2}) is extremely accurate in the whole parameter space of the problem for the external image. For the internal image $L_2$ is also very accurate in the region where it has an arc shape. This is the relevant region for the arc cross section as it is the only configuration where the image can have a large $L/W$ in the tangential direction. We found that $W_c$ (Eq.~\ref{Wc}) is a good 
approximation to the width of the image, 
except perhaps close to the Einstein ring. Using these two choices the axial ratio takes a very simple form
\begin{equation}
\frac{L}{W}=\frac{L_2}{W_c} = \left(\frac{1}{R_0}\arcsin \left(\frac{R_0}{s}\right)\pm 1\right).
\label{L2Wc}
\end{equation}
This expression shows explicitly the existence of a maximum value for the length-to-width ratio
given by $\left(L/W\right)_{\rm max} =\pi/\left(2R_0\right)\pm 1$, which corresponds to the formation of an Einstein ring ($s=R_0$). This value is easy to understand, as the two images are touching at their extrema on the verge to form the ring, such that the maximum value for $L$ can be approximated by $\pi$, and $\left(L/W\right)_{\rm max} \simeq \pi/(2R_0)$. 
From the expression (\ref{L2Wc}) above it is easy to find the threshold value $s_{\rm th}$ such that $L/W \ge R$. For $s \le R_0$ an Einstein ring is formed and this region does not contribute to the arc cross section.

\begin{figure*}
	\includegraphics[width=\textwidth]{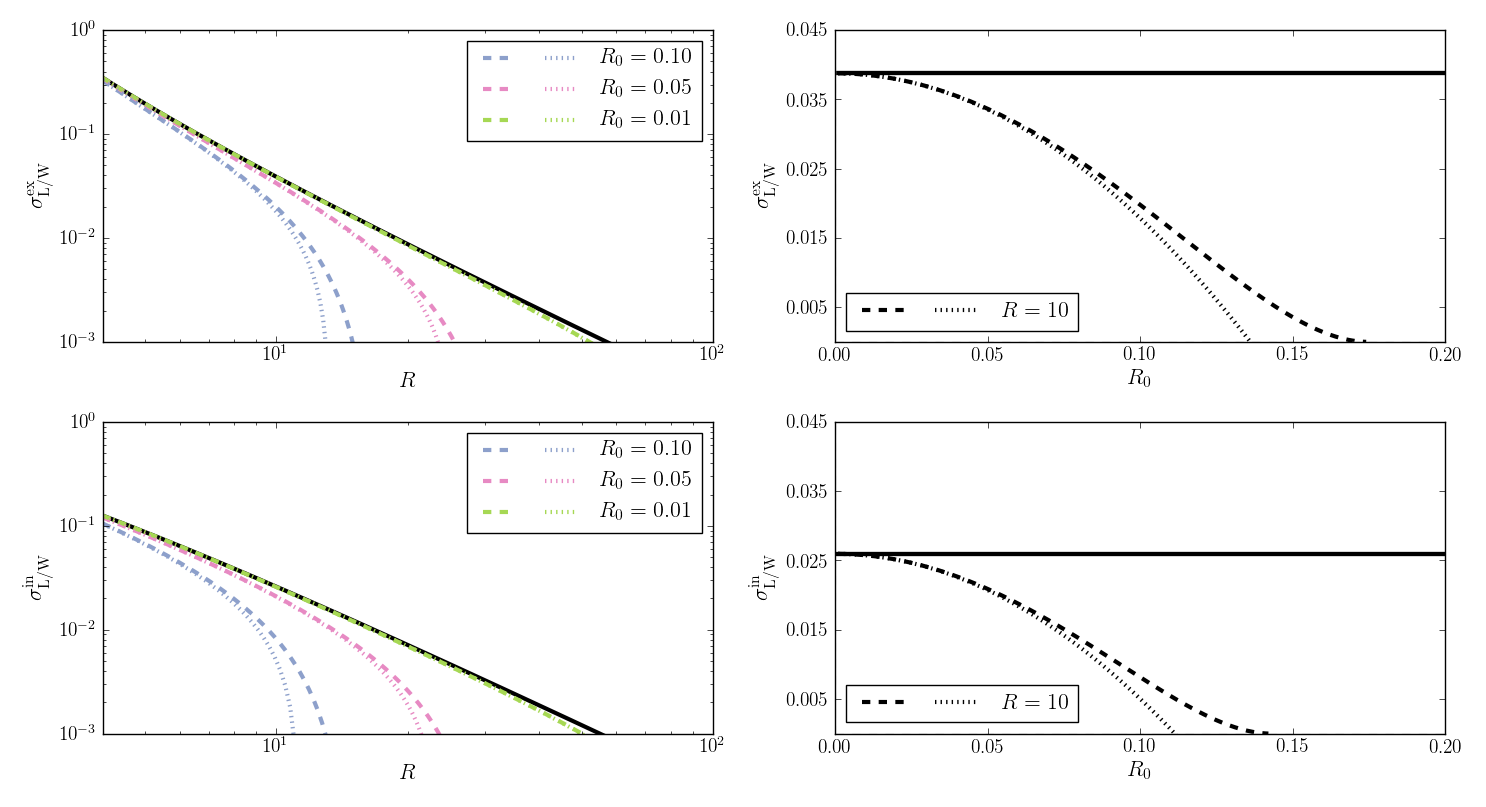}
    \caption{ Arc formation cross sections (dashed curves), perturbed arc formation cross sections (dotted curves) and infinitesimal approximation (black solid line) for fixed value of the source radius (left) and for a fixed value of the threshold (right). The first and second rows correspond to the external and internal arc, respectively.}
    \label{fig:cross1}
\end{figure*}

We can compute the cross section for the formation of each arc (internal and external) individually. The total cross section will simply be the sum of the two cross sections (as in Eq.~\ref{eq:sigmaRinf}). If the two arcs have length-to-width ratios above the threshold, that source position will count twice for the total cross section, if only one arc satisfies this condition, it will be counted once. Below we show the results for the individual arc cross sections, which are determined by $R_0 \le s \le s_{\rm th}$ and are given by
\begin{equation}
\sigma_{L/W} = \pi R_0^2\left(\csc^2\left(R_0 (R \mp 1) \right)-1\right).
\label{sigmaLW}
\end{equation}
This cross section is shown in fig. \ref{fig:cross1} (dashed line), along with the cross section for infinitesimal sources (Eq.~\ref{eq:sigmainex}, solid line). We see that the cross section for finite sources goes to zero for $R \ge \left(L/W\right)_{\rm max}$ as no arcs can be formed with length-to-width ratio above this value, as pointed out in \citet{Rozo2008}, in contrast to the infinitesimal source case.

For small values of $R_0$ (and far from the Einstein ring formation) we may use the expression~(\ref{LWapprox}) for 
$L/W$. Solving the third order equation for $s_{\rm th}$ for a given $R$, and taking the lowest order in $R_0$ leads to the cross section
\begin{equation}
\sigma_{L/W}^P = \frac{\pi}{\left(R \mp 1\right)^2} - \frac{2}{3}\pi R_0^2,
\label{sigmaWLP}
\end{equation}
where again, we have subtracted the region where Einstein rings are formed ($\pi R_0^2$). This expression shows the first order correction for finite sources to cross section for infinitesimal circular sources (Eq.~\ref{eq:sigmainex}).
Notice that, although the finite source $L/W$ (Eq.~\ref{LWapprox}) is higher than the infinitesimal one (Eq.~\ref{eq:mag}), there is a lower limit in $s$ such that Einstein rings are formed. We are excluding this region from the cross section, whereas this effect is not present for infinitesimal sources.

In Fig.~\ref{fig:cross1} we show the perturbative cross section as a function of $R$ and $R_0$, along with the complete cross section  (Eq.~\ref{sigmaLW}) and the infinitesimal one (Eq.~\ref{eq:mag}).
We see that the first order correction for finite source size is a very good approximation for lower values of the length-to-width threshold 
and captures the behavior of the full cross section until close to the formation of the Einstein ring.

\section{Summary}
\label{sec:summary} 

In this paper we investigated the geometrical properties of the images of finite circular sources lensed by a SIS, aiming to compute the magnification and arc cross sections. First we obtained their area $\mathcal{A}$, length $L$, and width $W$, testing several expressions for the latter two. The area is written in terms of elliptic integrals covering all possible source 
positions (Eqs. \ref{eq:areaexact} and \ref{eq:Aring}). We found that the length $L_4$ (Eq.~\ref{l4_def}), which is currently the most commonly adopted  to measure arcs, is virtually indistinguishable from the exact integration along the arc ridgeline. The alternative definition $L_2$ (Eq.~\ref{L2}) also provides an excellent approximation to better than 1\% precision for the external arc 
and has a simple analytical form. The width $W_i$ defined from the area  and length is in good agreement with the width at the arc center and with the mean width $\overline{W}$, as long as the right correction factor is applied in the later case. For the SIS the correction factor proposed in \citet{2012A&A...547A..66R} is manifest.
We obtain perturbative expressions to the lowest nontrivial order in $R_0$
for the area and all length and width definitions, which show clearly the finite source corrections to the solutions for infinitesimal sources.

From these quantities, we derive the total magnification $\mu_{\mathcal{A}}$ and the length-to-width ratio $L/W$ of the arcs.
We obtain an approximate solution for $\mu_{\mathcal{A}}$ well inside the Einstein ring regime. Far from this regime, both 
$\mu_{\mathcal{A}}$ and $L/W$ are enhanced for finite sources with respect to the infinitesimal case. The length-to-width ratio 
is expressed in a very simple form for $L_2/W_c$ (Eq. \ref{L2Wc}), 
which is a good approximation until the onset of Einstein ring formation.

Finally we apply these results to derive the cross sections. For  
the magnification, $\sigma_{\mu}$, we obtain the exact solution 
from the numerical inversion of the elliptical function, which is valid for sources of arbitrary size (as long as their radius is smaller than the Einstein one), 
including both arcs and rings.
We also obtain an approximation in simple form, 
valid for all magnification thresholds (Eqs.~\ref{sigmagsmall} and \ref{sigmaEring}). 
For the arc cross section, $\sigma_{L/W}$, we obtain a solution in 
terms of elementary functions (Eq.~\ref{sigmaLW}), for a specific choice of the length and width definitions,
valid for all length-to-width thresholds until the formation of an Einstein ring. We also derive a perturbative solution (Eq.~\ref{sigmaWLP}) 
showing explicitly the finite source correction.
We show that the cross sections vanish for thresholds above a given value ($R_{\max} \simeq \pi/(2 R_0)$, $\mu_{\max} = 4/R_0$), which is a behavior also noted in simulations \citep[see e.g.,][]{Bartelmann,Rozo2008,2011ApJ...734...52H,2012ApJ...755...46L}. This is easy to understand as $\mu_{\mathcal{A}}$ and $L/W$ are bounded in the case of finite sources.

In Appendix~\ref{sec:arcellipse} we compare the geometrical properties of the SIS arcs with those from the ArcEllipse. 
The results justify the use of the $\mathcal{A}$--$L$--$W$ and
$\overline{W}$--$W$ relations valid for ellipses to the case of gravitational arcs (at least those from SIS and circular sources), as has been done in previous works using simulations \citep[e.g.,][]{Bartelmann, 2002ApJ...573...51O, 2012A&A...547A..66R}.

In Appendix~\ref{sec:MagxR} we discuss a formalism by \citet{Fedeli} to include finite source effects in the cross section computation, showing that it yields a good approximation to the results of this paper 
for small sources.

\section{Concluding Remarks}
\label{sec:Conclusion}

We have presented a first study of the magnification and arc cross sections 
as computed from the exact solution for the images of finite sources.
The choice of a simple lens and source model allowed us to work all expressions up to the cross section in analytical form. Despite the simplifying assumptions of SIS lens and circular sources, this example is  
not of
purely pedagogical 
interest. Indeed, this combination of models has successfully been used to reproduce the observed abundance of sub-millimeter sources 
\citep{2012ApJ...755...46L}.
Furthermore our approach clarifies the results obtained empirically using ray-tracing simulations, such as the scaling of the maximum magnification $\mu_{\max}$ with the source size \citep{2012ApJ...755...46L}. It becomes clear why the magnification cross section for finite sources is enhanced  
for moderate magnifications and has a cutoff for $\mu > \mu_{\max}$. This behavior is also seen in the ray-tracing results \citep[e.g. compare Fig.~3 of][with our Fig.~\ref{fig:sigtot}]{2011ApJ...734...52H}.

Analytical solutions for the magnification of finite sources have been obtained in the literature \citep[e.g.][]{1964MNRAS.128..295R,2005ApJ...634...77I,Dobler2006}. However, to the best of our knowledge, these results have not been used previously to obtain the magnification cross section and its applications. Also, we are not aware of analytic studies on the length-to-width ratio of arcs, except the approximative method of \citet{Fedeli}.
Finite source effects in the magnification and arc cross sections have been studied through ray-tracing techniques \citep[e.g.,][]{2002ApJ...573...51O,2011ApJ...734...52H,2012ApJ...755...46L}. When applicable, our results are in agreement with these studies and provide a clear interpretation of some finite source effects addressed by them.

The approach laid out in this paper paves the way for similar studies using  more generic lens and source models. 
For example, some results can be readily extended for elliptical sources and SIS lenses (de Freitas et al, in prep.). This approach can be applied to other known analytical solutions for arcs from singular isothermal models, including external shear \citep{Dobler2006},
elliptical mass distributions \citep{Dobler2008}, 
and the combination of the two including source ellipticity (D\'umet-Montoya, et al., in prep.).
Furthermore, the method can be applied to lenses with more generic mass distributions using the analytic solutions for arcs from the perturbative approach of 
\citet[][see also \citet{2008MNRAS.390..945P}]{Alard2007}. 
Even if these cases do not lead to analytic expressions all the way down to the 
cross sections, the approach employed in this paper can speed up numerical computations by orders of magnitude as compared to ray-tracing methods \citep{Dobler2008} and provide hindsight on the solutions, and we expect it to be employed in realistic applications of arc statistics.

\section*{Acknowledgements}

V. P. de Freitas is funded by the Rio de Janeiro State Research Foundation (FAPERJ, E-26/200.279/2015). M.~Makler is partially supported by the National Council for Scientific and Technological Development -- CNPq 
and FAPERJ. Fora Temer. 
MM acknowledges the hospitality of Fermilab, where part of this work was done.



\bibliographystyle{mnras}
\bibliography{sisreferences} 



\appendix
\section{Gravitational arcs and the ArcEllipse shape}
\label{sec:arcellipse}Appendix~\ref{sec:arcellipse}

The ArcEllipse \citep{Furlanetto} is a simple geometrical figure to represent arc shapes. It is constructed by distorting an ellipse, such that its major axis is bent into an arc of a circle.
Therefore, instead of keeping constant the weighted squared sum of the distances to the Cartesian axes of coordinates, as in a standard ellipse, the ArcEllipse considers the distances perpendicular and tangential to a circle.
The ArcEllipse is thus the set of points whose distances from a point on the circumference along the tangential direction ($r_c\Delta\theta$) and along the radial direction ($\Delta r$) satisfy 
\begin{equation}
\left(\dfrac{r_c\Delta\theta}{a}\right)^2 + \left(\dfrac{\Delta r}{b}\right)^2 = 1,
\end{equation}
where $r_c$ is the radius of curvature of the circle (and of the constructed arc), $a$ is the length along the circle and $b$ is the width at the center in the radial direction (akin to the semi-axes of an ellipse).

Choosing the curvature center to coincide with the center of the polar coordinates we have $\Delta r = x-r_c$ and $\Delta\theta = \phi -\tilde{\theta}$, where $\tilde{\theta}$ is the orientation of the ArcEllipse center. Solving the quadratic expression above, we have
\begin{equation}
\label{eq:arcellipseeq}
x^{(\pm)} = r_c \pm b\sqrt{1-\left[\dfrac{r_c(\tilde{\theta}-\phi)}{a}\right]^2},
\end{equation}
where $x^{(+)}$ and $x^{(-)}$ delimit the inner and outer boundaries of the ArcEllipse, respectively. 

The extremities of the arc occur when $x^{(+)}=x^{(-)}$, similarly to the SIS case, 
and are given by $\phi_i = \tilde{\theta}-a/r_c$ and $\phi_f = \tilde{\theta}+a/r_c$. 

Eq.~(\ref{eq:arcellipseeq}) is akin to expression (\ref{eq:xmaismenosstheta}), except that now the ridgeline $\left(x^{(+)}+x^{(-)}\right)/2$
is, by construction, a segment of a circle and the curvature center is the center of the coordinate system. Therefore, all length definitions from $L_2$ to $L_6$ (Eqs.~\ref{L2}--\ref{L6}) coincide for the ArcEllipse and are given by
\begin{equation}
L_{AE} = r_c\left(\phi_f-\phi_i\right) = 2a.
\end{equation}
The width at the center of the arc is given as in Eq. (\ref{Wc}):
\begin{equation}
W_{AE} = \left(x^{(+)} - x^{(-)}\right)\bigg| _{\phi=\tilde{\theta}} = 2b.
\end{equation}
Therefore, the ratio $L/W$ is given by $L_{AE}/W_{AE} = a/b$, exactly as in the ellipse case. 

The area is computed in a similar fashion as in Eq.~(\ref{area}) and is given by \citep{Furlanetto}
\begin{equation}
\label{eq:Aae}
\mathcal{A}_{AE} = \dfrac{\pi}{4}L_{AE}W_{AE},
\end{equation}
which is identical to the area of an ellipse with semi-axes $a$ and $b$.

\begin{figure*}
\centering
\includegraphics[width = \textwidth]{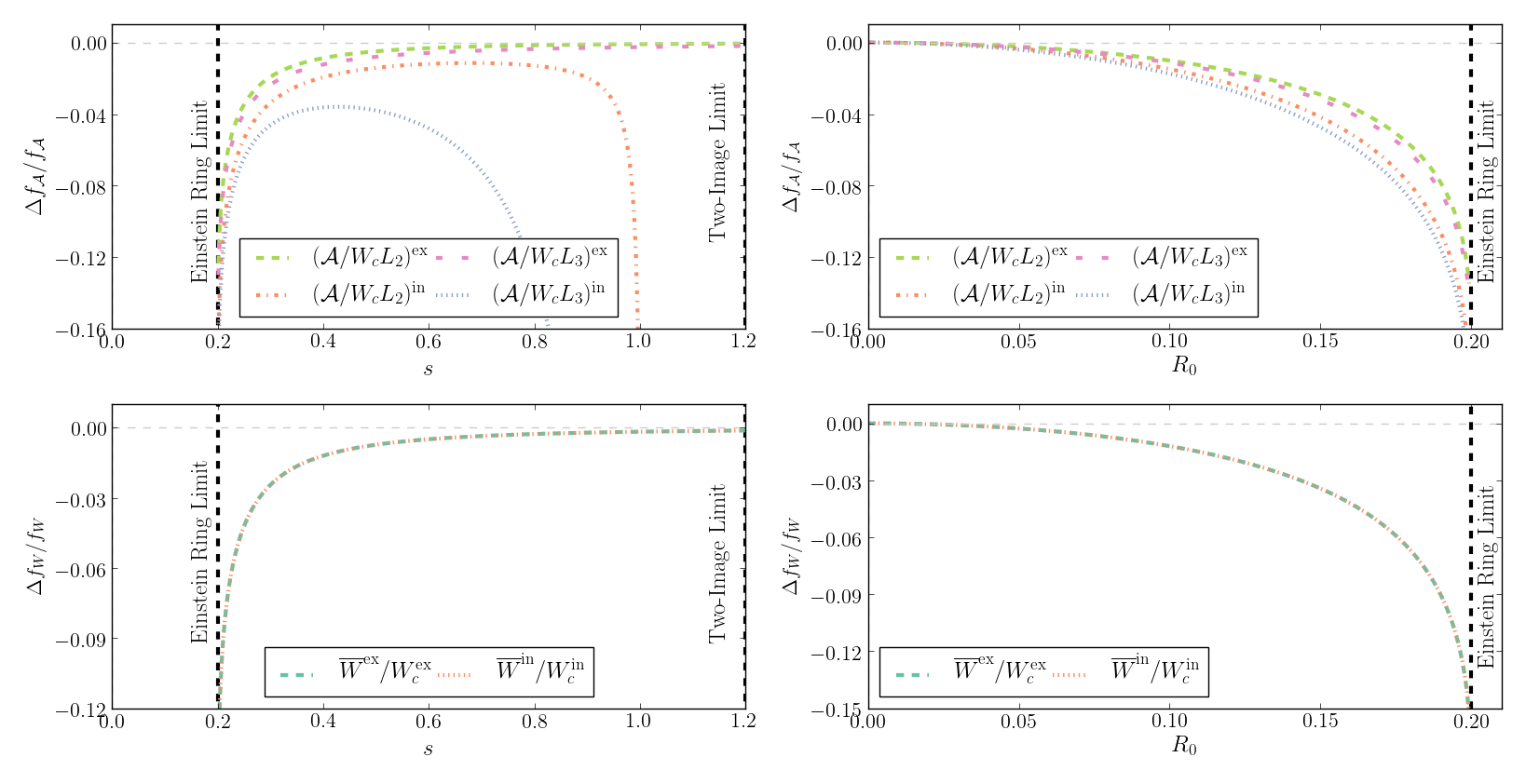}
\caption{Relative difference of the area form factor $f_\mathcal{A} = \mathcal{A}/LW$  (top panel) and 
the width form factor $f_W = \overline{W}/W_c$  (bottom panel) 
between the ArcEllipse and the images of circular sources lensed by a SIS. The parameters are $R_0=0.2$ (left) and $s = 0.2$ (right).}
\label{fig:arcellipsecomparison}
\end{figure*}

We may define the ratio $f_\mathcal{A} =\mathcal{A}/\left(LW\right)$ as a form factor of a geometrical figure. In the case of the ArcEllipse this factor is $\pi/4$. Of course, this factor will be smaller the sharper the extremities of the figure, i.e., when the width decreases significantly away from the center.
For an annulus segment, for example, $f_\mathcal{A} =1$.
In Fig.~\ref{fig:arcellipsecomparison} (upper panels) we show the relative difference of the area form factors for the ArcEllipse and for the arcs produced by a SIS with circular sources $\Delta f_\mathcal{A}/f_\mathcal{A} = \left(f_\mathcal{A}^{SIS}-f_\mathcal{A}^{AE}\right)/f_\mathcal{A}^{AE}$.
We consider two definitions of $L$ for making the comparison with the SIS case: the exact length along the arc ridgeline $L_3$ and the approximation $L_2$ that we used to compute the arc cross section.
We use the same width definition as for the ArcEllipse, i.e., the width at the center of the arc $W_c = 2R_0$.
For the external arc, the difference in the form factor is remarkably small until close to the formation of the Einstein ring. At that point the arc extremities become sharper and the shape deviates more substantially from the ArcEllipse. As expected, the difference between the two length definitions is negligible. 
For the internal image, in the regime in which it is arc shaped ($s \le 1/2$) the form factors are also very similar to the ArcEllipse, except close to the Einstein ring formation. 
In the whole range of $s$ the difference is smaller using $L_2$, 
especially when the image loses its arc shape and takes a droplet shape.
The difference in form factor decreases for smaller sources, i.e., the smaller the source the more the images look like ArcEllipses.

The mean width of the ArcEllipse along the radial direction is obtained as in Eq.~(\ref{meanW}) and is given by
\begin{equation}
\label{eq:Wae}
\overline{W}_{AE} =\dfrac{\pi}{2} b =\dfrac{\pi }{4}W_{AE}.
\end{equation}
We may define another form factor of a figure by the ratio $f_W = \overline{W}/W_c$. Again, arcs with thiner/sharper extremities will have lower values of $f_W$.
In Fig.~\ref{fig:arcellipsecomparison} (lower panels) we show the relative difference of the width geometrical factor for the ArcEllipse and the SIS arcs, $\Delta f_W/f_W = \left(f_W^{SIS}-f_W^{AE}\right)/f_W^{AE}$. In this case, the behavior is the same for the internal and external arcs.
We see that the differences are very small, except close to the Einstein ring limit. 
However, even when the two images are touching, the differences are $\lesssim 15\%$, as for $f_\mathcal{A}$.

A simple recipe to obtain an ArcEllipse that matches the image of circular source lensed by a SIS is: $i$) center the ArcEllipse at the curvature center of the image, from Eqs.~(\ref{centeroffsets})
and (\ref{newr0}), such that $r_c=r_0$, $ii$) choose the ArcEllipse length such that $a = r_c\theta_0$, with $\theta_0$ given by Eq.~(\ref{theta0}), $iii$) choose the ArcEllipse width such that $b = R_0$. The resulting figure is almost identical to the SIS arc, except close to the formation of an Einstein ring. 
In the case of the internal image its shape is very similar to an ArcEllipse in the regime in which the image is arc shaped.
The deviations from the ArcEllipse shape are well described by the differences in the form factors shown in Fig.~\ref{fig:arcellipsecomparison}.
The smaller the ratio $R_0/s$, the SIS arc solution is closer to the ArcEllipse. 
In brief, for most configurations, the ArcEllipse is an excellent representation to SIS arcs.


\section{Approximate Computation of finite source effects in the lens plane}
\label{sec:MagxR}

As mentioned in Sec. \ref{sec:Cross Section}, to obtain the cross sections for finite sources one has to obtain their images, 
which usually implies solving the lens equation numerically and is computationally expensive. The conditions $L/W> R$ or $\mu > \mu_{\rm th}$ then define the area in the source plane for which the images satisfy these conditions (Eq.~\ref{eq:sigdef}).
On the other hand, the computation for infinitesimal sources can be carried out on the lens plane, from Eq.~(\ref{eq:siglens}), without the need of inverting the lens equation, as the local magnification and axial ratios are naturally defined in this plane.

\citet{Fedeli} have proposed an approximate method to compute the cross section for finite sources in the lens plane.
According to their proposal, extended sources are taken into account by convolving
the eigenvalue ratio 
with a suitable window function quantifying the source size. In their method the axial ratio $L/W$ is approximated by 
\begin{equation}
h = \int_{\mathbb{R}^2} R_\lambda(y)g(y) {\rm d}^2y= \int_{\mathbb{R}^2}{R_\lambda(x) g(x) \frac{{\rm d}^2x}{|\mu(x)|}},
\label{hR}
\end{equation}
where $R_\lambda$ is the axial ratio for infinitesimal circular sources, as defined as in Eq.~(\ref{eq:Rlambda}), and $g(y)$ is a window function representing the surface brightness distribution of the source. For a uniform circular source $g(x) = g(y(x))$ is zero outside the image of the source 
and takes the value $1/\left(\pi R_0^2\right)$ inside.

In \citet{Fedeli} the expression for $L/W$ is further approximated so as to avoid explicitly carrying out the integral (Eq.~\ref{hR}). By assuming that the eigenvalues of the mapping do not change significantly across a single source, they find an approximate solution for $h$ in terms of derivatives of $R_\lambda$ (Eq. \ref{eq:Rlambda}) and the mean values of the eigenvalues $\lambda_{r,t}$ (Eq. \ref{eq:lambdas}) (see their expression A.16). These authors apply their formalism to lenses with Navarro---Frenk---White \citep{nfw1} profiles and elliptical potentials to represent merging clusters \citep{Fedeli}. They find a good agreement between these approximations and ray-tracing simulations, but with a computation time reduced by a factor of $\sim 30$ with respect to the latter. 

In the case of the SIS, it is simple to obtain an analytical expression for $h$ without the need of any further approximation. In this case we have $R_\lambda(x) =\mu(x)$, such that the integral (\ref{hR}) is simply $ \mathcal{A}^{\rm ex,in}/\left(\pi R_0^2\right)$, which is exactly the magnification of each image (Eq.~\ref{eq:areaexact}).
Therefore, in this approximation, the arc cross section is obtained by replacing $L/W$ with $\mu_{\mathcal{A}}^{\rm ex,in}$.
This cross section, which will be referred to as 
$\sigma_h$,
is computed as in section \ref{sec:sigmaLW} applying the condition $\mu_{\mathcal{A}}^{\rm ex,in} \ge R$ (and excluding the arc formation region $s < R_0$). In this case the cross section does not have an analytic solution and has to be obtained numerically from the inversion of the elliptic integral. 

\begin{figure*}
	\includegraphics[width=\textwidth]{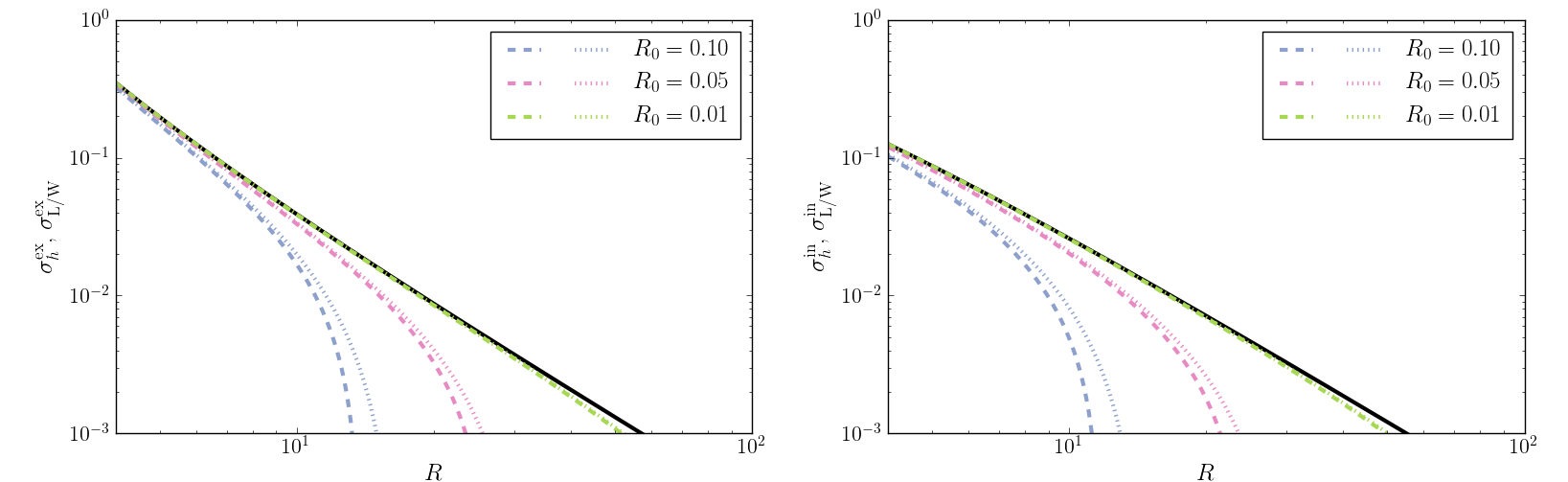}
   \caption{Comparison between the arc cross sections using $L/W$ for finite sources (dotted), the convolution $h$ of the local eigenvalue ratio on the image (dashed), and the infinitesimal source approximation (black solid line)  for fixed values of the source radius. Left: external arc. Right: internal arc.}
    \label{fig:sis_cross_magcomp}
\end{figure*}

In Fig.~\ref{fig:sis_cross_magcomp} we show the resulting $\sigma_h$
as a function of $R$ for a few values of $R_0$, together with the cross section $\sigma_{L/W}$ 
(Eq.~\ref{sigmaLW}).
We see that replacing the axial ratio $L/W$ by $h$ does capture the dependence of the finite source cross section with $R_0$ and $R$ and provides a good approximation to the arc cross section for $R \lesssim 10$. The approximate cross section is systematically lower than $\sigma_{L/W}$, which is qualitatively consistent with the results of \citet{Fedeli}, where the proposed approximation appears to underestimate the cross section as compared to the ray-tracing simulations (see their Fig. 1).

We may obtain an explicit expression for $\sigma_h$ for $R_0~\ll~1$ by using the perturbative expansion of the magnification for each arc (Eq.~\ref{eq:areaaprox}) and following the same procedure as in section \ref{sec:sigmaLW}, which gives
\begin{equation}
\sigma_h^P = \frac{\pi}{\left(\mu_{th} \mp 1\right)^2} - \frac{3}{4}\pi R_0^2.
\label{sigmamuP}
\end{equation}

\begin{figure*}
	\includegraphics[width=\textwidth]{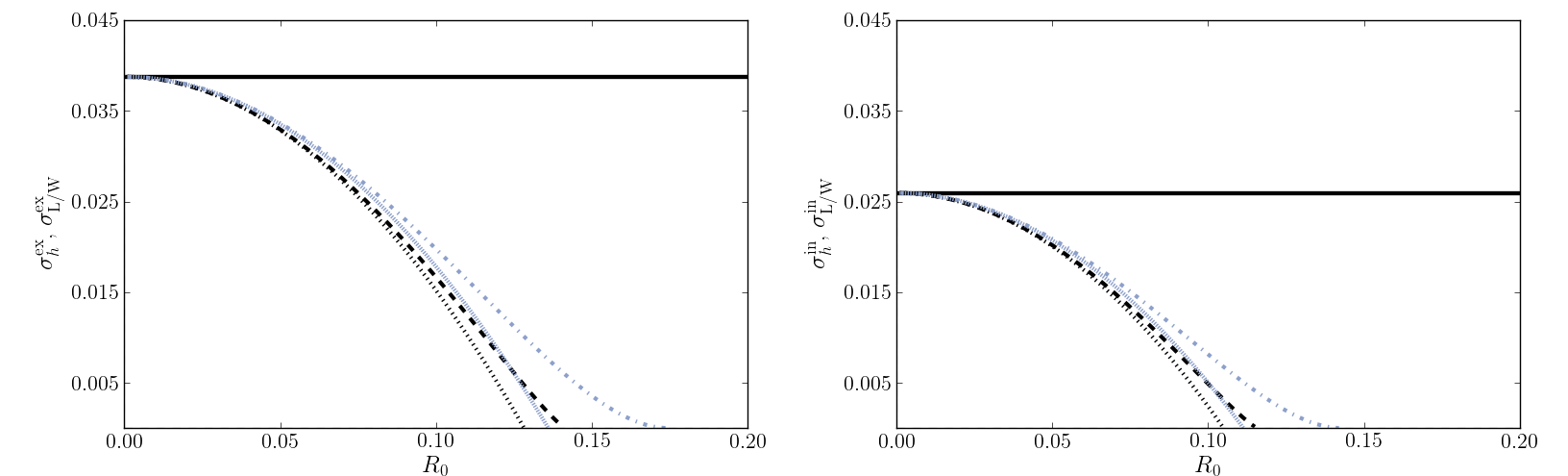}
    \caption{Arc cross sections using $L/W$, $\sigma_{L/W}$ (dash-dotted curve) and $\sigma^P_{L/W}$ (densely dotted), using the approximation $h$, $\sigma_h$ (dashed curve) and $\sigma_h^P$ (dotted line) and the infinitesimal source approximation (black solid line) for $R = 10$. 
    Left: external arc. Right: internal arc.}
    \label{fig:cross2}
\end{figure*}

This result is shown in Fig.~\ref{fig:cross2}, for $R=10$, along with $\sigma_h$, $\sigma_{L/W}$, its perturbative expansion in $R_0$ (Eq.~\ref{sigmaWLP}) and the infinitesimal cross section. As expected all four expressions for finite sources are similar for $R_0\ll1$.

By comparing Eqs.~(\ref{sigmaWLP}) and (\ref{sigmamuP})
we see that the first correction terms to the  cross section for finite sources differ
by about $10\%$ from using the exact $L/W$ and $h$. Therefore, the difference between the two cross sections is less than $10\%$ at the perturbative level.

We conclude that the approximation based on equation (\ref{hR}) is accurate, at least in the case of a SIS lens and circular sources.

\label{lastpage}
\end{document}